\documentclass{aa}  
\usepackage{array}
\newcolumntype{L}[1]{>{\raggedright\let\newline\\\arraybackslash\hspace{0pt}}m{#1}}
\newcolumntype{C}[1]{>{\centering\let\newline\\\arraybackslash\hspace{0pt}}m{#1}}
\newcolumntype{R}[1]{>{\raggedleft\let\newline\\\arraybackslash\hspace{0pt}}m{#1}}
\usepackage{graphicx}
\usepackage{natbib}
\bibpunct{(}{)}{;}{a}{}{,}
\usepackage{stfloats}
\usepackage{supertabular}
\usepackage{tabularx}
\usepackage{color}
\usepackage{caption}
\usepackage{subcaption}
\usepackage{txfonts}
\usepackage{amsmath}
\usepackage{amssymb}
\usepackage{multirow}
\usepackage{hyperref}
\usepackage{longtable,lscape}
\usepackage{rotating}

\usepackage[normalem]{ulem}

\begin{document} 

   \title{An XMM--Newton catalogue of BL Lacs\thanks{Tables B.1, B.2, B.3, B.4 and C.1 are only available in electronic form at the CDS via anonymous ftp to cdsarc.u-strasbg.fr (130.79.128.5) or via http://cdsweb.u-strasbg.fr/cgi-bin/qcat?J/A+A/}}


   \author{I. de la Calle P\'erez\inst{1},
           N. \'Alvarez Crespo\inst{2},
           E. Racero\inst{3} \& 
           A. Rouco\inst{4} 
          }

   \institute{Quasar Science Resource S.L. for the European Space Agency (ESA), European Space Astronomy Centre (ESAC), Camino Bajo del Castillo s/n, 28692 Villanueva de la Ca\~{n}ada, Madrid, Spain
              \email{icalle@sciops.esa.int}
    \and
    European Space Agency (ESA), European Space Astronomy Centre (ESAC), Camino Bajo del Castillo s/n, 28692 Villanueva de la Ca\~{n}ada, Madrid, Spain
    \and
    Serco Gestión de Negocios SLU for the European Space Agency (ESA), European Space Astronomy Centre (ESAC), Camino Bajo del Castillo s/n, 28692 Villanueva de la Ca\~{n}ada, Madrid, Spain
    \and
    Center for Interdisciplinary Exploration and Research in Astrophysics (CIERA) and Department of Physics and Astronomy, Northwestern University, 1800 Sherman, Evanston, IL, 60201, USA}


\abstract
   {}
   {We present an {\it XMM--Newton} catalogue of BL Lac X-ray,
   optical, and UV properties based on cross-correlation with the
   1151 BL Lacs listed in the fifth edition of the Roma--BZCAT.}
   {We searched for the X-ray counterparts to these objects in the field of view of all pointed observations in the {\it
     XMM--Newton} archive over nearly 20 years of mission. The
   cross-correlation yields a total of 310 {\it XMM--Newton} fields which
   correspond to 103 different BL Lacs.  We homogeneously analysed data from the three EPIC cameras (X-ray) and OM (optical/UV) using the {\it
     XMM--Newton} SAS software, and produced images, light curves, and spectral products for BL Lacs detected in any of the three EPIC cameras. We tested two
   different phenomenological models, {\it log parabola} and {\it power law},
   with different variations of the absorbing column density and  extracted
   their parameters. We derived time-variability information from
   the light curves following well-established statistical methods and
   quantified variability through statistical indicators.  OM magnitudes and
   fluxes were computed wherever possible.}
   { We see that the {\it log parabola} model is
     preferred over the {\it power law} model for sources showing higher
     fluxes, which might indicate that curvature is intrinsic to BL Lacs and
     is only seen when the flux is high. We present the results of our analysis as a catalogue of X-ray spectral properties of the sample in
     the 0.2 - 10 keV energy band as well as in the optical/UV
     band. We complete the catalogue with multi-wavelength information at radio and $\gamma$-ray energies.  }
   {}


   \maketitle
%

\section{Introduction}
\label{sec:intro}

\par According to the unified scheme of active galactic nuclei (AGNs), a
blazar is considered to be a radio-loud AGN that displays highly variable,
beamed, non-thermal emission covering a broad wavelength range from radio
to $\gamma$-ray energies \citep{Urry1995}. The properties of the blazar
non-thermal emission suggest a relativistic origin in a jet oriented at
a small angle to our line of sight \citep{1978PhyS...17..265B}. The blazar
class encompasses BL Lacertae (BL Lac) objects and flat-spectrum radio quasars
(FSRQs). The most striking difference between the two  lies in their optical spectra, with the former showing
 weak or no emission or absorption lines \citep[$EW <$ 5
  $\AA$,][]{1991ApJ...374..431S} and the latter strong broad emission
lines. One difficulty in the study of BL Lacs is indeed that in many cases
it is not possible to measure any redshift because of a lack of features in
{\bf their} optical spectra \citep[see e.g.][]{2016AJ....151...95A}. Blazars, and in
particular BL Lacs, are very strong $\gamma$-ray emitters, being the most numerous extragalactic sources at high energies
\citep[see][]{2015ApJS..218...23A,2012ApJS..199...31N,2010ApJS..188..405A}.

\par Observationally, the spectral energy distribution (SED) of blazars in a
$\nu$F$_\nu$ representation shows two broad distinctive peaks \citep[see  e.g.][]{Padovani1995}. The most widely accepted interpretation of the first bump of the SED is that it is due to synchrotron radiation of relativistic electrons
moving along the jet. Different models offer alternative solutions as to the origin of the second
spectral component, namely leptonic or hadronic. In the leptonic model, the
second peak is the result of inverse Compton (IC) scattering of low-energy
photons to high energies by the very same electrons that produce the
synchrotron peak. The up-scattered photons can be the synchrotron photons themselves \citep[self-synchrotron Compton model,
  SSC;][]{Ghisellini1999,1992MNRAS.258..776G} or ambient photons of different
origins \citep[external Compton model, EC; see review][]{Sikora2001}.  On the
other hand, hadronic models establish the origin of the second peak as
synchrotron emission from high-energy protons or photoproduction induced by
accelerated protons \citep[see e.g.][]{Mannheim1996,Aharonian2000,Mucke2003}.
Blazars presenting the first peak in their SED at UV/X-ray energies
($\nu_{peak}^{S} > $ 10$^{15}$ Hz) are referred to as high-synchrotron peaked
blazars (HSPs). For intermediate-synchrotron peaked blazars (ISPs), the
synchrotron peak frequency lies at 10$^{14}$ Hz $< \nu_{peak}^{S} <$ 10$^{15}$
Hz.  Those blazars presenting their synchrotron peak frequency at
$\nu_{peak}^{S} < $ 10$^{14}$ Hz, between radio and optical wavelengths, are
known as low-synchrotron peaked blazars
\citep[LSPs;][]{2010ApJ...716...30A}. BL Lacs can belong to all these three
categories, while most FSRQs have been classified as LSPs.

\par \cite{Fossati1998} showed that the SED in blazars is correlated with the
bolometric luminosity, forming the so-called blazar sequence. As the
bolometric luminosity increases, blazars become redder, that is, their two bumps
in the SED show smaller peak frequencies and the Compton peak becomes
increasingly dominant. \cite{1998MNRAS.301..451G} interpret this as a result
of the different amounts of radiative cooling suffered by electrons in
different sources.  On the other hand, \cite{2012MNRAS.420.2899G} consider
that this blazar sequence is most likely a selection effect due to the
selection of sources from radio- and X-ray-flux-limited samples. Using Monte
Carlo simulations to populate the blazar sequence, these latter authors consider that
BL Lacs could not be plotted at high luminosity and high $\nu_{peak}^{S}$ ; not
because these objects do not exist but rather because it is not possible to
measure their redshifts and therefore it is not possible to estimate their luminosities.
\cite{2017MNRAS.469..255G} used the Third Large-Area telescope AGN Catalogue
\citep[3LAC,][]{2015ApJ...810...14A} to revisit the blazar sequence. They
constructed their average SED using $\gamma$-ray {\bf luminosities}, finding an
anti-correlation with the synchrotron peak frequency. Their new results support
the blazar sequence scenario.


The {\it Fermi Gamma-ray Space Telescope} launched in 2008 has revealed
more than 5000 sources.  With a sky density of $\sim$0.1 sources deg$^{2}$,
blazars and particularly BL Lacs dominate the $\gamma$-ray sky and constitute
more than 75\% of the associated sources in all releases of the Fermi
catalogues \cite[1FGL,][]{2010ApJS..188..405A},
\cite[2FGL,][]{2012ApJS..199...31N}, \cite[3FGL,][]{2015ApJS..218...23A} and
\cite[4FGL,][]{2020ApJS..247...33A}.  However, all these catalogues present a
fraction of about one-third of the sources as unassociated or unknown. The
discovery that blazars occupy a narrow region in the WISE IR colour-colour
space, the so-called WISE blazar strip \citep[WBS,
][]{2011ApJ...740L..48M,2012ApJ...748...68D}, has been used to search for
blazar-like sources within the positional uncertainty regions of the
unidentified or unassociated $\gamma$-ray sources (UGSs) leading to many new blazar candidates
\citep{2014ApJS..215...14D,2019ApJS..242....4D,2016AJ....151...95A,2020ApJS..248...23D}.

\par Several studies have proven that BL Lacs exhibit high amplitude and rapid
variability on timescales that can vary from hours to months, and so the nature of
the X-ray spectra in BL Lacs is variable and shows a complex behaviour
\citep[see
  e.g.][]{2006ApJ...651..782Z,2004ApJ...601..165F,2005A&A...443..397B}.  The
shape of the X-ray spectrum can provide information that can be used to reveal the emission components, as X-ray emission probably originates in the inner
parts of the relativistic jet.  The transition in the SED between synchrotron
and IC in BL Lacs is located with the X-rays, and therefore characterising the
spectra in this band where both processes can contribute to the X-rays is of
special relevance. When the mechanism responsible for the emission is
synchrotron, we observe a steep power-law energy distribution ($\alpha > 1$,
$F \propto \nu^{-\alpha}$) due to the tail of its spectrum, while if IC is the
dominant process we observe a flat component ($\alpha < 1$). In the SED of
HSPs, the X-ray radiation comes from the high-energy end of the synchrotron,
whereas in LSPs it comes from Compton scattering.

\par Here we present a catalogue that compiles 310 
observations corresponding to 103 different BL Lacs observed with {\it
  XMM--Newton} over nearly 20 years of operations. We make use of the fifth
edition of the catalogue Roma-BZCAT \cite[`BZCAT'
  hereafter,][]{2015Ap&SS.357...75M}, which constitutes the most comprehensive
list of blazars known to date and is based on multi-frequency surveys and
information reported in the literature. BZCAT contains 3561 sources, of which
1151 are confirmed BL Lacs.  Here we report those BL Lacs observed by {\it
  XMM--Newton} and reported in the {\it } archival public
observations\footnote{See \url{http://nxsa.esac.esa.int}} up to September
2019. We fit six different models to each {\it XMM--Newton} source where
sufficient counts are available to produce a spectrum: two different
phenomenological models, {\it log parabola} and {\it power law}, each with
three different flavours for the absorbing column density. We study their
parameters and choose the best-fit model. The catalogue includes light curves,
derived variability parameters, and optical and ultraviolet (UV) information
when available.
 
\par The paper is organised as follows. In Sections~\ref{sec:sample_selection}
and~\ref{sec:data_analysis} we describe the sample selection with observations and data
analysis,  respectively. In Section~\ref{sec:the_catalogue}, we
describe the catalogue, the products and how to access it. In
Section~\ref{sec:results} we present and discuss the spectral properties and the results of the best-fit model, and in Section~\ref{sec:conclusions} we present our conclusions.

\par Unless otherwise stated, we assume a flat $\Lambda$CDM cosmology with a
Hubble constant $H_0 =$ 70 km s$^{-1}$ Mpc $^{-1}$, total matter density
$\Omega_m =$ 0.27 and dark energy density $\Omega_\Lambda =$ 0.73. Magnitudes
throughout this paper refer to the AB magnitude system
\citep{1974ApJS...27...21O}.

\
\section{Sample selection}
\label{sec:sample_selection}

\begin{figure*}[h]
  \centering
  \includegraphics[width=.95\textwidth]{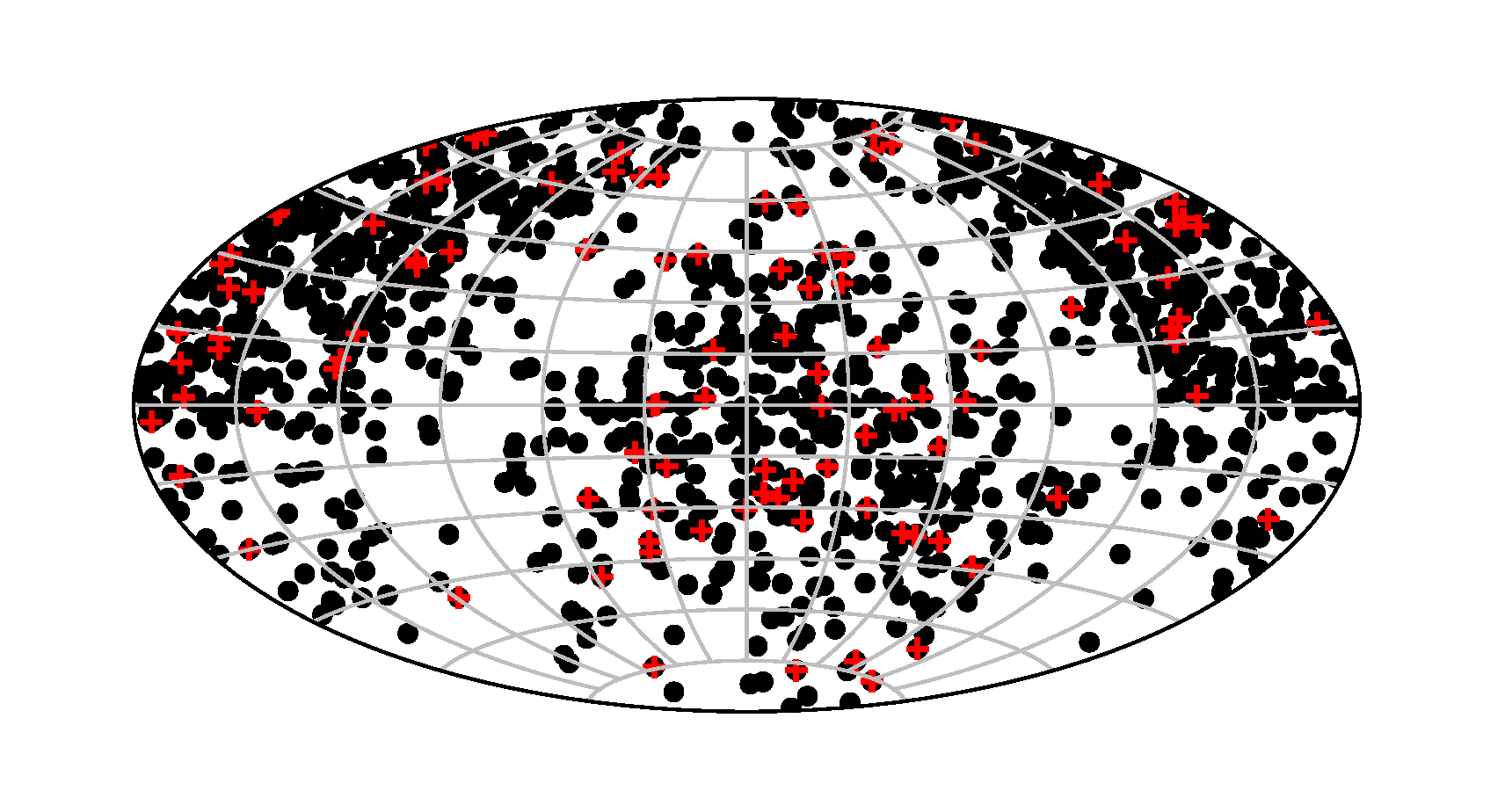}
  \caption{ Equatorial coordinates of the
   BL Lacs from  BZCAT, indicated as black points. Red crosses mark the distribution of those sources 
   contained in the {\it catalogue of XMM--Newton BL Lacs}.}
  \label{sample_XMMVeron}
\end{figure*}

\par The sample we present here is the result of a cross-correlation between the
1151 BL Lac subsample given in the fifth version of the Roma--BZCAT and all the
pointed science public observations available in the {\it XMM--Newton} archive. Overall, observations covered by the catalogue include
the period from {\bf March 2000 to September 2019}. Calibration observations cover
up to September 2020. Their cross-correlation with the {\it XMM--Newton}
archive yields a total of 310 {\it XMM--Newton} observations that contain a
BZCAT source, corresponding to 103 different {\bf BZCAT} sources. The 310 {\it XMM--Newton}
observations are selected by searching in squares of sizes
$S_x=0.5^\circ$/cos(Dec), $S_y=0.5^\circ$, centred at the positions of the
BZCAT sources to see if the centre of an {\it XMM--Newton} observation is contained within this box, which is equivalent to saying that the BZCAT source is
contained within any field of view of an {\it XMM--Newton}
observation. Figure \ref{sample_XMMVeron} shows the sky distribution in
equatorial coordinates of the BZCAT BL Lacs and those present in our {\it  XMM--Newton BL Lac Catalogue}. Because of the selection procedure, the sample
is heterogeneous, certainly not complete, and sources are observed in different
epochs.

\par In the {\it XMM--Newton} BL Lac catalogue, {\bf 101 of the 103} different sources are
detected in the EPIC cameras (EPIC-MOS and EPIC-pn, respectively) as described
in Section~\ref{sec:data_analysis}. The BL Lac 5BZBJ2214+0020 (1/103) is
detected in one pointing and undetected in another. For the remaining (2/103)
BL Lacs, flux upper limits are provided. The BL Lacs detected in our catalogue
are a combination of {\it XMM--Newton} targets (41/101) and serendipitous
objects (60/101) found anywhere within the field of view of the EPIC
cameras. Within the 41 sources that are targets, 4 sources are the target of
observations in some observations and field sources in other observations. Any source lying more than 0.5$\arcmin$ from
the nominal {\it XMM--Newton} pointing is considered serendipitous. All calibration sources are considered as targets, even if
displaced from the nominal pointing,.

\par We classify the BL Lacs according to the estimated value of the
synchrotron peak frequency $\nu_{peak}^{S}$ calculated in the observed frame
as defined in Section~\ref{sec:intro}, following the nomenclature used in the
Fourth Large-Area Telescope AGN Catalogue
\citep[4LAC,][]{2020ApJ...892..105A}.  For some BL Lacs these values are
reported in the 4LAC, while for the rest of the sources we build their SEDs
source by source using the {\it ASDC SED Builder Tool}\footnote{See
  \url{https://tools.ssdc.asi.it/SED}}, which collects all available data in
the literature. We estimate the peak of the synchrotron emission
$\nu_{peak}^{S}$ using a third-degree polynomial fit on the low-energy hump:

\begin{equation}
   log \ (\nu F_{\nu})=A(log \ \nu)^{2} + B (log \ \nu) +C
,\end{equation}

\noindent where the error associated with $log \ \nu_{peak}^{S}$ is typically
$\sim$0.5 due to the use of non-simultaneous broadband data
\citep{2013AJ....145...31K}.

With this criteria, about 48\% of the BL Lacs in the {\it XMM--Newton}
catalogue are classified as HSPs, 26\% as ISPs, and 21\% as LSPs. For the remaining
fraction of sources, it is not possible to make a classification because of a
lack of sufficient broadband data to perform an adequate fit.

\section{Observations and data analysis}
\label{sec:data_analysis}

\subsection{Instrument and observations}
\label{subsec:instobs}
\par The {\it XMM--Newton} observatory carries several coaligned X-ray
instruments: the European Photon Imaging Camera (EPIC) and two reflection grating spectrometers \citep[RGS1 and RGS2,][]{Jansen2001}. The EPIC cameras
consist of two metal-oxide semiconductors \citep[EPIC-MOS,][]{Turner2001} and
one pn junction \citep[EPIC-pn,][]{Struder2001} CCD array, which have a
$\sim$30$\arcmin$ field of view (FOV) and can offer 5 - 6{\arcsec} spatial
resolution and 70 - 80 eV energy resolution in the 0.2 - 10 keV energy
band. {\it XMM--Newton} also has a co-aligned  optical/UV
telescope of 30 cm in diameter (Optical Monitor, OM), providing strictly simultaneous observations
with the X-ray telescopes \citep{Mason2001}.  It has three optical and three
UV filters, with effective wavelengths V: 543~nm, B: 450~nm, U: 344~nm, UVW1:
291~nm, UVM2: 231~nm, and UVW2: 212~nm covering a 17 $\times$ 17 arcmin$^{2}$
FOV (but with the actual imaged sky area being dependent on the user-chosen
mode) and a point spread function (PSF) of less than 2{\arcsec} FWHM
---depending on filter--- over the full FOV \citep[]{Rosen2020}.

\par
\begin{figure}
  \centering
  \includegraphics[width=.45\textwidth]{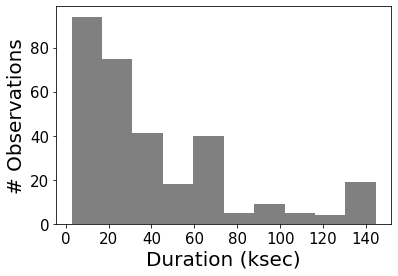}
  \caption{Distribution of the exposure time (in ks) of single observations included in the {\it catalogue of XMM--Newton BL Lacs}.}
  \label{fig:histksec}
\end{figure}

 The information on the 310 {\it XMM--Newton} observations reported in this
work comes mainly from the EPIC detector. EPIC observations present in this
catalogue have been taken under timing and imaging mode, this latter in
combination with different window modes (Small Window, Large Window, and Full
Frame) and filters (Thick, Medium, and Thin). EPIC data are available for all
observations in at least one of the three cameras. For 35\% (36/103) of the
objects, several observations exist. Observing times per observation range between 2.7~ks and 144~ks (see Fig.~\ref{fig:histksec}), and
observing times per object range between 2.7~ks and 2.2~Ms, spanning several years 
in some cases.

\par Data provided by the OM exist for 45\% (46/103) of the objects in at
least one of the filters. For 8 objects, data exists in all six filters
(three optical and three UV). For 12 objects, there are data in all three
optical filters, and for 13 there are data in all three UV filters. For 26
objects there are data in at list one optical and one UV filter. For 5
objects, no OM data are available due to the presence of bright objects in the
field of view. 

\par
The following sections describe the EPIC data-reduction procedure.

\subsection{Data reduction}
\label{sec:data_reduction}

\par Our sample has been uniformly analysed using the {\it XMM--Newton}
standard Science Analysis System \citep[SAS, v18.0.0;][]{Gabriel2004} and most
updated calibration files. Different observations of the same source are not
combined for any purposes, including source detection. If several exposures exist for any given
observation of a source, then the one with the longest
exposure time is used. Event lists are produced for the three EPIC cameras
following the standard SAS reduction procedure. Periods of high-background
activity are removed following the standard method \citep{Lumb2002}, that is,
monitoring the rate of events with pattern `0' and energies $>$10~keV (the
limits that define the good time intervals (GTIs) are defined by background
rates EPIC-pn$_{rate}<1.0~cts/sec$ and EPIC-MOS$_{rate}<0.8~cts/sec$).

\par EPIC-MOS1 exposures taken in TIMING mode contained in our data sample are
discarded and not used in our analysis. Details explaining why this is the case can be found
in the {\it EPIC Status of Calibration and Data Analysis}; latest update Oct
2019 \citep{Smith2019}.

\par
{\bf We use the standard {\tt omichain} SAS task to analyse OM data. The processing chain performs source
detection and computes rates and instrumental magnitudes as well as fluxes for
every source detected.}

\par In the following sections we highlight the main analysis steps needed
to go from EPIC source detections (~\ref{subsec:src_detection}), through
coordinate corrections (~\ref{sec:src_posrec}), and source cross-match
(~\ref{sec:src_crossmatch}) to end with the extraction of the catalogue
products (~\ref{sec:src_products}).

\subsubsection{Source detection}
\label{subsec:src_detection}

\par The source-detection algorithm is applied only to exposures taken in
IMAGING mode. If the exposure time is less than 100~secs in any given EPIC
exposure after GTI filtering, the exposure is not considered for source
detection and the source is flagged as not detected. All the TIMING exposures
in our data sample contain a source, so they are flagged as detections by default.

\par The GTI-filtered event-detection criteria have to be established for the
EPIC cameras. For this purpose, images with a bin size of 40 pixels/bin are
produced in five different energy ranges (EPIC-pn: 300 - 500, 500 - 2000, 2000 -
4500, 4500 - 7500, 7500 - 12000 eV; EPIC-MOS: 200 - 500, 500 - 2000, 2000 -
4500, 4500 - 7500, 7500 - 12000 eV). These images are used by the SAS task
{\tt edetect\_chain} to establish source detection. This task is run
simultaneously over the five energy bands and each camera independently. A
positive detection requires a detection likelihood threshold of 10.
A radial statistical error is associated to every detected source,
computed as $\sqrt{RA_{err}^2 + DEC_{err}^2}$, where RA$_{err}$ and
DEC$_{err}$ are the 1$\sigma$ statistical uncertainties in the derived RA and
DEC coordinates respectively.

\subsubsection{Source position rectification}\label{sec:src_posrec}

\par Subsets of objects within the {\it XMM--Newton} FOV, extracted from large
astrometric reference catalogues, are used to rectify the source positions
derived by {\tt edetect\_chain}. Before this is done, the SAS task {\tt
  srcmatch} is used to create a unique EPIC source list using information from
all EPIC cameras in which the source has been detected. This unique list
contains average source information, such as fluxes. Within this unique list,
the associated RA and DEC are an average of those found independently for each
camera, and the positional error is a combination of statistical and systematic
errors. For reference, the average 1$\sigma$ positional error for the whole
Fourth {\it XMM--Newton} Serendipitous Source Catalogue
\citep[4XMM,][]{Webb_2020} is better than $\sim$ 1.7{\arcsec}, with a
standard deviation of 1.4. As for the systematic errors, the relative
astrometry within each camera is accurate to within 1.5{\arcsec} over the full
FOV.

\par The SAS task {\tt catcorr} is run over the unique EPIC source list to
update the sky coordinates by cross-matching them with up to three reference
external catalogues, (i) the USNO B1 catalogue \citep{2003AJ....125..984M}, (ii)
the 2MASS catalogue \citep{2006AJ....131.1163S}, and (iii) the SDSS (DR9)
catalogue \citep{2012ApJS..203...21A}, to find optical or infrared (IR)
counterparts and to find the small frame shifts or rotations that optimise
the match.  If sufficient matches are found within the FOV to optimise the
correction, these shifts are then applied to the positions of all the EPIC
sources in the field. The average 1$\sigma$ field correction is of the order
of 2.5{\arcsec}. Where {\tt catcorr} fails to obtain a statistically reliable
result, the associated systematic error is assumed to be 1.5{\arcsec}.

\par The average (over all EPIC cameras) and external catalogue rectified sky
coordinates are used as the coordinates of the {\it XMM--Newton} X-ray sources
that make up our catalogue. The catalogue includes the coordinates {\tt
  XMM\_RA\_NOCORR}, {\tt XMM\_DEC\_NOCORR,} and {\tt XMM\_RADEC\_ERR\_NOCORR}
before astrometric rectification, and {\tt XMM\_RA}, {\tt XMM\_DEC,} and {\tt
  XMM\_RADEC\_ERR} after astrometric rectification.  Two further columns in
the catalogue, {\tt AstCorr} and {\tt PosCorr,} identify if astrometric
rectification has been applied and if the fitting was successful in providing
updated coordinates.

\subsubsection{Catalogue cross-match}
\label{sec:src_crossmatch}

\par Once we have the {\it XMM--Newton} sources with rectified positions, a
positive match with a BZCAT source is considered if:

  \begin{equation}
    \frac{distance}{\sqrt{\sigma_{BZCAT}^2 + \sigma_{XMM}^2}} \le 2\sigma  ,\end{equation}

\noindent where {\it distance} refers to the distance between the {\it
  XMM--Newton} and BZCAT source in units of arcsec; $\sigma_{BZCAT}$ is the
reported positional uncertainty of the order of 1{\arcsec}; and $\sigma_{XMM}$
the estimated {\it XMM--Newton} positional uncertainty, a combination of
$\sigma_{XMM}^2$ = $(\sigma_{STAT}^2 + \sigma_{SYS}^2)$, the statistical and
systematic uncertainty in the {\it XMM--Newton} position, where $\sigma_{STAT}$
= 1.5{\arcsec} on average, and $\sigma_{SYS}$ $\sim$ 1.5{\arcsec}. This translates to an average distance no greater than $\sim$ 5 {\arcsec}
(2$\sigma$) between the {\it XMM--Newton} detected source and the BZCAT source
in order to consider the X-ray source a match of the BZCAT one.

\par In summary, if the {\it XMM--Newton} detected source is within
$\sim$5{\arcsec} of the BZCAT one, we consider the X-ray source to be a
positive match. With this criteria, the cross-match yields 307/310 {\it
  XMM--Newton} observations with an identified BZCAT source, corresponding to
103 different BZCAT sources.

\subsubsection{Catalogue product extraction}
\label{sec:src_products}

\par Once a positive identification of the X-ray source has been established,
source and background extraction regions are defined to derive source light
curves and spectra.

\par For IMAGING mode, the source region is chosen to be circular around the
source centroid with a minimum radius of 20{\arcsec} and limited to a maximum
of 40{\arcsec}, whereas the background region is an annulus around the source
region with inner radius r$_{min}$ = 60{\arcsec} and outer radius r$_{max}$ =
120{\arcsec}. This selection ensures that the source region is greater than
the PSF of the instrument (4 - 6{\arcsec}) and that the background region is
away from any possible source contamination.  Additional X-ray sources in the
background region are masked out by removing those with a detection likelihood
greater than 50. The SAS task {\tt eregionanalyse} is used to maximise the S/N
calculated from the source counts, the encircled energy fraction, and the
background counts to give the optimum source radius and source centroid. If
{\tt eregionanalyse} cannot find an optimal extraction radius, the exposure is
discarded by our analysis.

\begin{figure*}
  \centering
  \includegraphics[width=.9\textwidth]{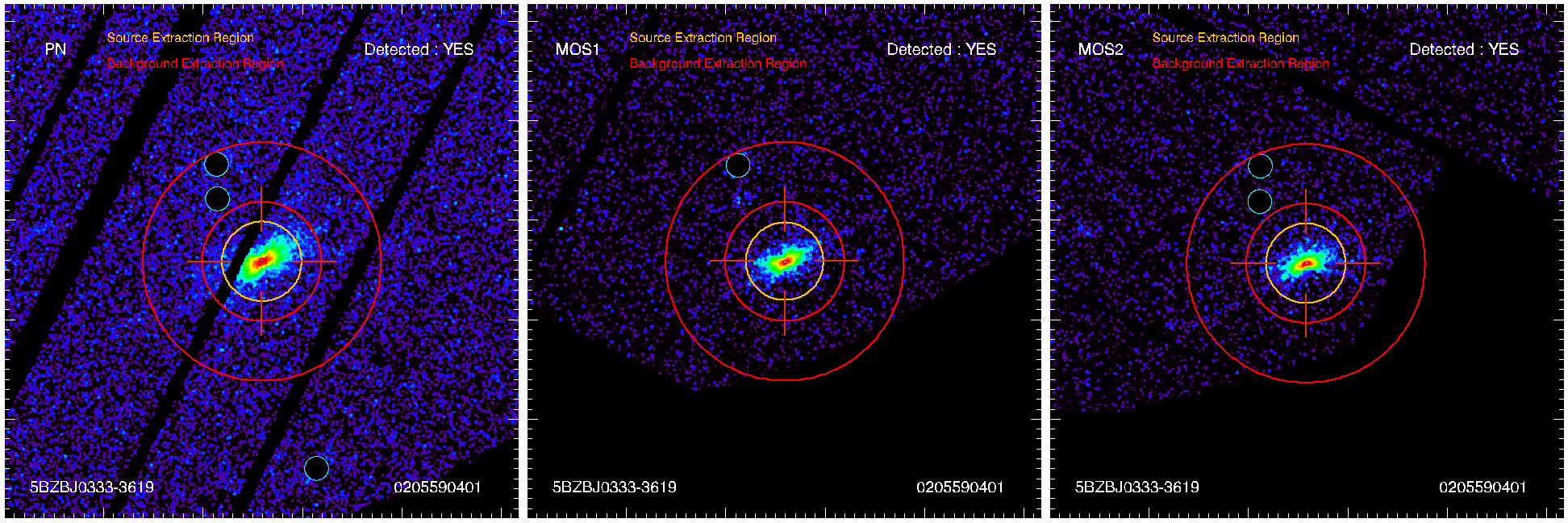}\\
  \includegraphics[width=.9\textwidth]{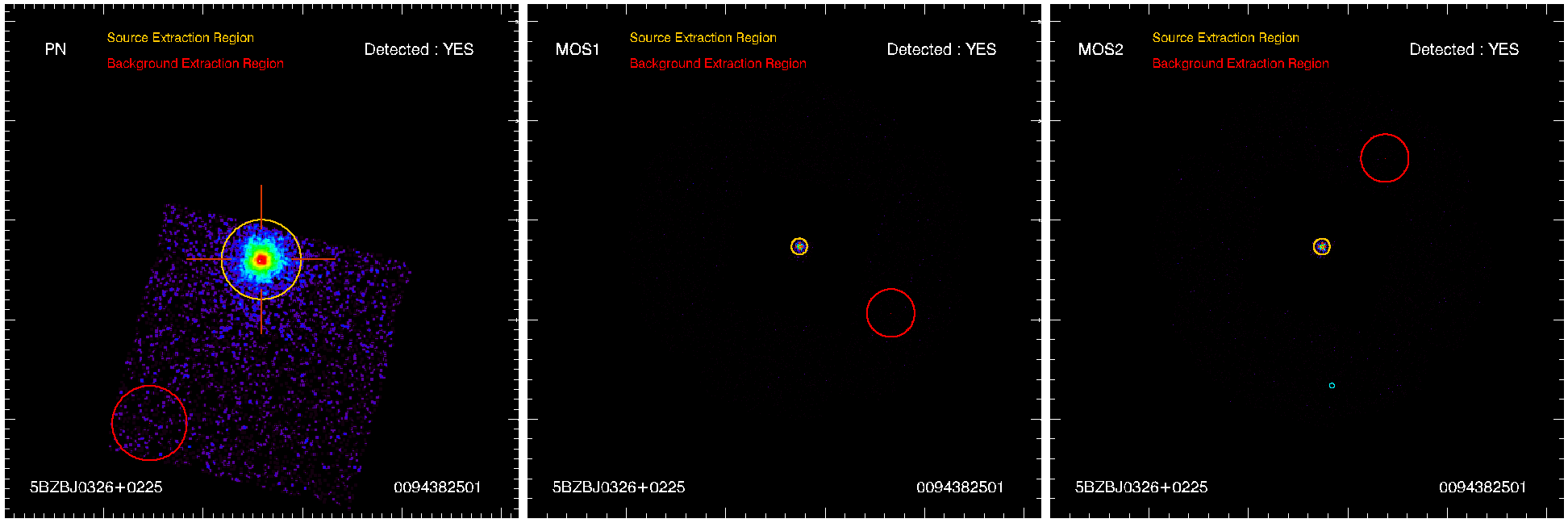}\\
  \caption{{\it Top:} EPIC IMAGING Full Frame mode source (orange) and background (red)
  extraction region definition for the source 5BZBJ0333-3619. Left panel shows PN, middle panel shows 
  MOS1, and right panel MOS2 cameras. {\it Bottom:} EPIC IMAGING Small Window mode 
  source (orange) and background (red) extraction region definition for the source 5BZBJ0326+0225. Left panel shows PN, middle panel shows 
  MOS1, and right panel MOS2 cameras.}
  \label{epic_regions}
\end{figure*}

\par An exception to the above rule for IMAGING mode exposures is the
definition of background extraction regions taken from exposures in Small Window mode. To avoid contamination from the bright central source, for
EPIC-pn the background region is taken from a circular region from a corner of
the available window rather than from an annulus around the source. The
circular background-extraction region has a radius of 37.5{\arcsec}. For
EPIC-MOS, the background region is taken from a circular region from one of the
outer CCDs, in particular from CCD5. The radius of the background-extraction
region is 120{\arcsec}. Figure~\ref{epic_regions} illustrates how extraction
regions are selected for the two IMAGING mode EPIC exposures, Full Frame and
Small Window.

\par For TIMING mode exposures in EPIC-pn, both source and background are extracted from boxes in RAWX versus RAWY. The source region is defined from a box
of 16 $\times$ 199 RAW X,Y units with centre $X_{centre} = <RAWX>$ and
$Y_{centre}$ = 101. The background is extracted from a box of 16 $\times$199
RAW X,Y units with centre $X_{centre} = 8.5$ and $Y_{centre}$ = 101. The
background box is moved along RAWX from the edge of the detector by 0.5 units.

\par For TIMING mode exposures in EPIC-MOS, the source and background are extracted from boxes in RAWX versus TIME. The source region is defined from a box
of 60 RAWX units and full time (Exposure) interval with centre $X_{centre} =
<RAWX>$ and $Y_{centre} = T\_Start+<Exposure>/2.$. The background is extracted
from a box of 20 RAWX units with centre $X_{centre} = 264.0$ and $Y_{centre} =
T\_Start+<Exposure>/2$, where $T\_Start$ is the exposure start time.

\subsubsection{Pile-up evaluation and correction}
\label{subsec:pileup}

\par For IMAGING mode exposures, the presence of `pile-up' is evaluated. To test
for pile-up we derived the source+background count rate extracted from a
circle around the source of 60{\arcsec} (this radius contains about 95\% of
the PSF). To establish the presence of pile-up, these count rates are compared
against pile-up limits given in Table 2 of \cite{Jethwa2015} which are in line with the following
criteria: a conservative limit where 2\% - 3\% flux loss and less than 1\% spectral distortion is incurred, and a tolerant limit allowing for 4\% - 6\%
flux loss and 1\% - 1.5\% spectral distortion. These limits are a function of
the EPIC observing submodes and are available for Full Frame, Large Window, and
Small Window modes, both for EPIC-pn and EPIC-MOS. If the submode is outside
these values, a tag {\tt UNK} is used to mark in the catalogue the presence of
pile-up. The number of IMAGING mode exposures in the catalogue where pile-up
is present is: 82/238 for EPIC-pn, 102/253 for EPIC-MOS1, and 114/284 for EPIC-MOS2.

\par An attempt is made to reduce the pile-up only below the 2\%-3\% flux loss
limit. This is only done to extract spectral products, the correction is not
applied to extract light-curve products. The procedure involves 
iteratively removing circles around the source centroid, starting with a radius of 10{\arcsec} and taking increasing steps of 5{\arcsec} until the pile-up is
below the 2\% - 3\% flux loss limit. In every step, the presence of pile-up is
tested up to a maximum removal radius of 30{\arcsec} (about 90\% of the
PSF). If at this point pile-up is still present, no more corrections are
attempted as no PSF would be left to extract any meaningful spectral
products. However, rather than throwing out the exposure, spectral products
are still derived from the full source region, and so care has to be taken in interpreting the spectral results derived from these exposures. Overall, the
pile-up correction worked for 54/82 EPIC-pn exposures, 56/102 EPIC-MOS1
exposures, and 67/114 EPIC-MOS2 exposures where pile-up was present.
Table~\ref{table:pileup_exp} collects the list of 53 observations where there
is at least one exposure where pile-up could not be removed. Of these,
25 observations contain no exposures free of pile-up, where 24 of these 25
correspond to observations of MRK~421 (5BZBJ1104+3812). In the remaining
28, it is possible to find at least one exposure that is free of pile-up. In
summary, out of the 310 observations contained in the catalogue, 25/310 suffer
from irrecoverable pile-up effects.

\begin{table*}[h!]
\centering
\caption{List of 53 observations with at
least one EPIC exposure where pile-up is present and cannot be corrected.}
\label{table:pileup_exp}
\begin{tabular}{lcccc} 
\hline\hline             
  BZCAT Object 
         & \multicolumn{1}{c}{Observation ID} 
         & \multicolumn{1}{c}{EPIC-pn} 
         & \multicolumn{1}{c}{EPIC-MOS1} 
         & \multicolumn{1}{c}{EPIC-MOS2} \\
\hline
5BZBJ1104+3812 & 0158971201 & NO & YES & YES  \\
               & 0136541001 & NO & YES & YES  \\
               & 0099280101 & NO & YES & YES  \\
               & 0658800101 & YES & YES & YES \\
               & 0136540901 & YES & YES & YES \\
               & 0158970101 & NO & NO & YES \\ 
               & 0411082701 & YES & YES & YES \\
               & 0560983301 & YES & YES & YES \\
               & 0411083201 & YES & YES & YES \\
               & 0656380801 & YES & YES & YES \\
               & 0658801301 & YES & YES & YES \\
               & 0099280201 & NO & YES & NO \\
               & 0153950701 & YES & YES & YES \\
               & 0136540801 & YES & YES & NO \\
               & 0510610201 & NO & YES & YES \\
               & 0411081901 & NO & YES & YES \\
               & 0560980101 & YES & YES & YES \\
               & 0136541201 & YES & YES & YES \\
               & 0658802301 & NO & YES & YES \\
               & 0411080301 & YES & YES & YES \\
               & 0099280301 & YES & YES & YES \\
               & 0153950601 & YES & NO & YES \\
               & 0136541101 & YES & YES & YES \\
               & 0150498701 & NO & YES & NO  \\
               & 0162960101 & YES & YES & YES \\
               & 0158971301 & NO & YES & YES \\
               & 0510610101 & NO & YES & YES \\
               & 0411081301 & YES & YES & YES \\
               & 0411081401 & YES & YES & YES \\
               & 0411081501 & YES & YES & YES \\
               & 0656380101 & YES & YES & YES \\
               & 0656381301 & YES & YES & YES \\
               & 0791780601 & YES & YES & YES \\
               & 0411081601 & YES & YES & YES \\
               & 0136540101 & YES & YES & YES \\
               & 0099280401 & YES & NO & NO  \\
               & 0136540701 & YES & YES & YES \\
               & 0810860201 & YES & YES & YES \\
               & 0658801801 & NO & YES & YES \\
5BZBJ1136+7009 & 0094170101 & YES & YES & YES \\
5BZBJ1221+3010 & 0111840101 & NO & NO & YES  \\
5BZBJ1428+4240 & 0300140101 & NO & YES & YES \\
               & 0111850201 & NO & NO & YES \\
5BZBJ1653+3945 & 0652570201 & NO & YES & YES \\
               & 0652570401 & NO & YES & YES \\
               & 0652570101 & NO & YES & YES \\
               & 0113060201 & NO & NO & YES  \\
               & 0113060401 & NO & NO & YES  \\
               & 0652570301 & NO & YES & YES \\
5BZBJ1959+6508 & 0850980101 & NO & YES & YES \\
5BZBJ2158-3013 & 0124930301 & NO & YES & NO  \\
5BZBJ2359-3037 & 0693500101 & NO & YES & YES \\
               & 0722860101 & NO & YES & NO  \\
 \hline
\end{tabular}\\
Column description. (1): Source name as given in BZCAT, (2): {\it XMM--Newton}
Observation ID, (3): flag indicating whether EPIC-pn is affected by pile-up,
(4): idem EPIC-MOS1 and (5): idem EPIC-MOS2.
\end{table*}

\subsubsection{Quality flags}
\label{quality}

\par Several quality flags are checked before catalogue products are
extracted. First, we check whether the GTI filtered exposure time is greater
than 100~s; if it is not, the analysis of the exposure stops here. If there
is a positive detection, the next step is to establish the presence of pile-up,
and if present, whether it can be corrected as described in
Section~\ref{subsec:pileup}. For the light-curve generation, a minimum of 500~s is required. Finally, the spectrum has to contain at least
300~counts to ensure a minimum quality spectral fitting.

\subsection{Catalogue products}\label{cat_products}

\par The catalogue products include EPIC images, light curves, and spectral
products. These are obtained from the GTI-filtered event lists as described in
Section~\ref{sec:data_reduction}. Light-curve and spectral products are
acquired from the source- and background-extraction regions as described in
Section~\ref{sec:src_products}. When available, OM images together with
magnitudes and fluxes are given too. All products have been visually screened
to check for problems. In the following sections, we describe how individual
products are obtained.

\subsubsection{Image generation}\label{src_products_image}

\par
\begin{figure*}
  \centering
  \includegraphics[width=.3\textwidth]{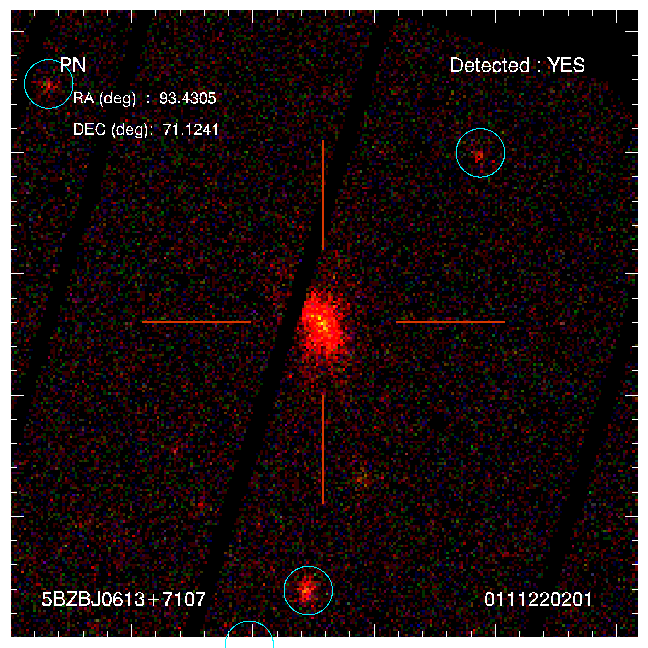}
  \includegraphics[width=.3\textwidth]{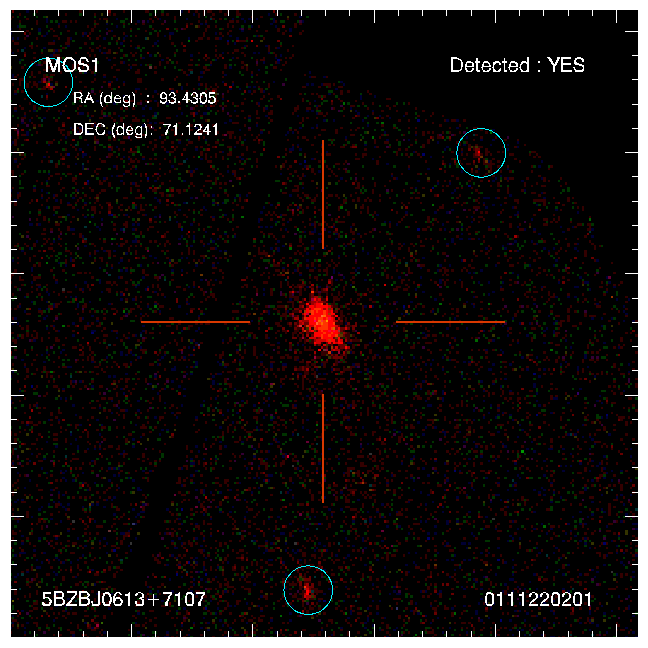}
  \includegraphics[width=.3\textwidth]{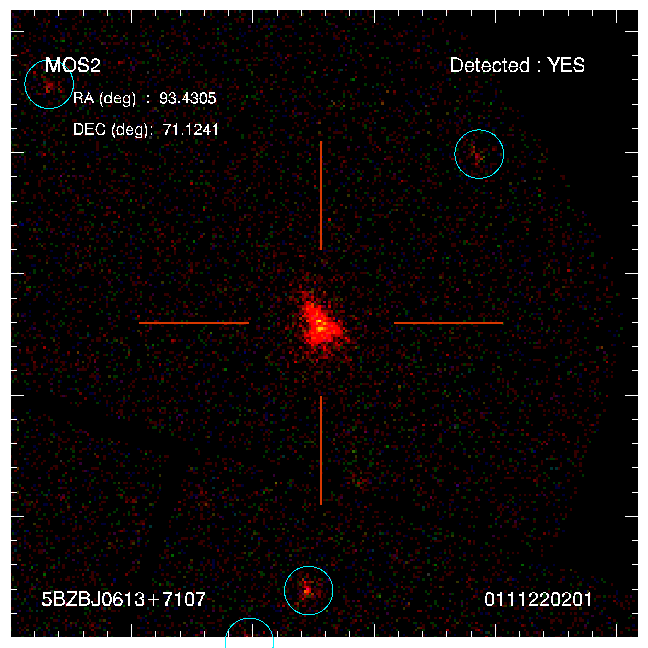}\\
  \caption{EPIC IMAGING mode thumbnail true colour images around the X-ray source 5BZBJ0613+7107 marked
            by a red cross (Events in red: 0.5 - 4.5 keV, green: 4.5 - 7.5 keV, blue: 7.5 - 12.0 keV). Blue circles 
            mark other X-ray sources in the field.}
  \label{epic_thumbnails}
\end{figure*}

 Images are produced for all individual IMAGING mode exposures in sky coordinates with a bin size of 40 units. These images are used for source
detection and are available in five energy ranges (EPIC-pn : 300 - 500, 500 -
2000, 2000 - 4500, 4500 - 7500, 7500 - 12000 eV; EPIC-MOS: 200-500, 500 -2000, 2000 - 4500, 4500 - 7500, 7500 - 12000 eV). Thumbnail true colour images
are produced around the X-ray source, as seen in Fig.~\ref{epic_thumbnails}
(Events in red: 0.5 - 4.5 keV, green: 4.5 - 7.5 keV, blue: 7.5 - 12.0 keV).

\subsubsection{OM image and source information}\label{om_products}

\par OM images and source information are obtained via the SAS task
     {\tt omichain} when OM exposures are available. This SAS perl script
     reduces OM Imaging mode data and produces individual mosaiced sky images
     and combined source lists with flux and magnitude information. For the
     purpose of this work, {\tt omichain} is run with the optional parameters
     {\tt processmosaicedimages} and {\tt usecat} set to true in order to
     maximise the astrometric accuracy of the detected sources and provide the
     deepest level of imaging product (mosaiced sky images) when possible.

\par {\tt omichain} comprises a number of SAS tasks. The imaging pipeline first processes the image for individual exposures, producing a set of OM
imaging-mode products including the list of detected sources per exposure through the {\tt omdetect} SAS task. In cases where more than one
exposure is available per filter, the {\tt ommosaic} SAS task generates a mosaic sky image per filter as a combination of the individual exposures. {\tt
  omsrclistcomb} then combines the source lists corresponding to the available
individual exposures. Finally, {\tt ommergelists} combines all the available
source lists for the different filters and produces an associated region
file. Source quality flags for OM are one of the outputs of {\tt omichain} and
are included in Table B.3 of the catalogue (see
Section~\ref{subsec:the_catalogue_description}).

\par The main OM information contained in our catalogue is  magnitudes and
fluxes in any of the six passbands where the source is detected. This
information is extracted from the observation source list produced by {\tt
  omichain}, which includes the results of the combination of multiple
exposures and different extraction apertures. These apertures are used to derive fluxes and are not fixed, but vary between 3" and 6". Fluxes provided
in our catalogue are not extinction corrected. For a more detailed discussion
on apertures used to extract fluxes and how to deal with extinction
corrections we refer the readers to \cite{Page2012}.

\par For the astrometric corrections of the detected sources, {\tt
  omsrclistcomb} identifies if a USNO catalogue file is available, and if so
corrects the astrometry ---depending on whether there are sufficient matches of
OM detections with USNO objects---in order to account for any offset between the OM and
USNO source adding the columns {\tt RA\_CORR} and {\tt DEC\_CORR} to the
combined master source list. This final list is then cross-matched with the
coordinates of our source of interest. We follow the same procedure as
described in \ref{sec:src_crossmatch} for the EPIC cameras. In this case, we consider $\sigma_{XMM\_OM}^2$ = $(\sigma_{STAT}^2 +
\sigma_{SYS}^2)$ {\bf the estimated positional uncertainty, a combination of} statistical and systematic uncertainties in the OM position, where
$\sigma_{STAT}$ ranges from 0.05 to 2.57{\arcsec}, with a mean position
uncertainty of 0.68{\arcsec} \citep{Page2012} and $\sigma_{SYS}$ $\sim$
0.7{\arcsec} \citep{Rosen2020}. In a similar manner as with EPIC sources, this
implies that a significant positive match is found if the BZCAT source lies
within $\sim$ 3{\arcsec} of the OM source. This is a purely geometrical
approach based on the statistical coordinates of the OM source and no attempt
has been made to use the extent of the OM source.  The observation source
lists contain information relative to the extension of the source. This
information is not contained in our catalogue.

\subsubsection{Light-curve generation}\label{src_products_lc}

\par Source and background light curves are produced in two energy ranges,
E$_{soft} = 0.2 - 2.0$ keV and E$_{hard} = 2.0 - 10.0$ keV, over bins of 500~s. For TIMING mode exposures, the low-energy limit of the {\it soft} band
light curve is set at 0.3~keV to avoid low-energy noise.

\par
\begin{figure*}[b]
  \centering
    \includegraphics[width=.46\textwidth]{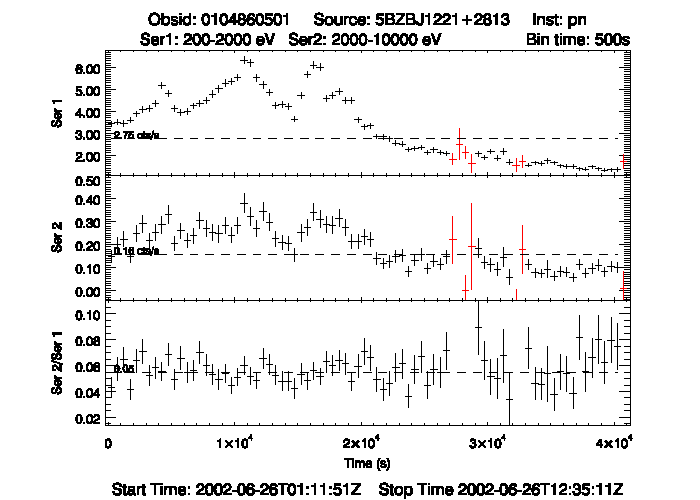}
    \includegraphics[width=.47\textwidth]{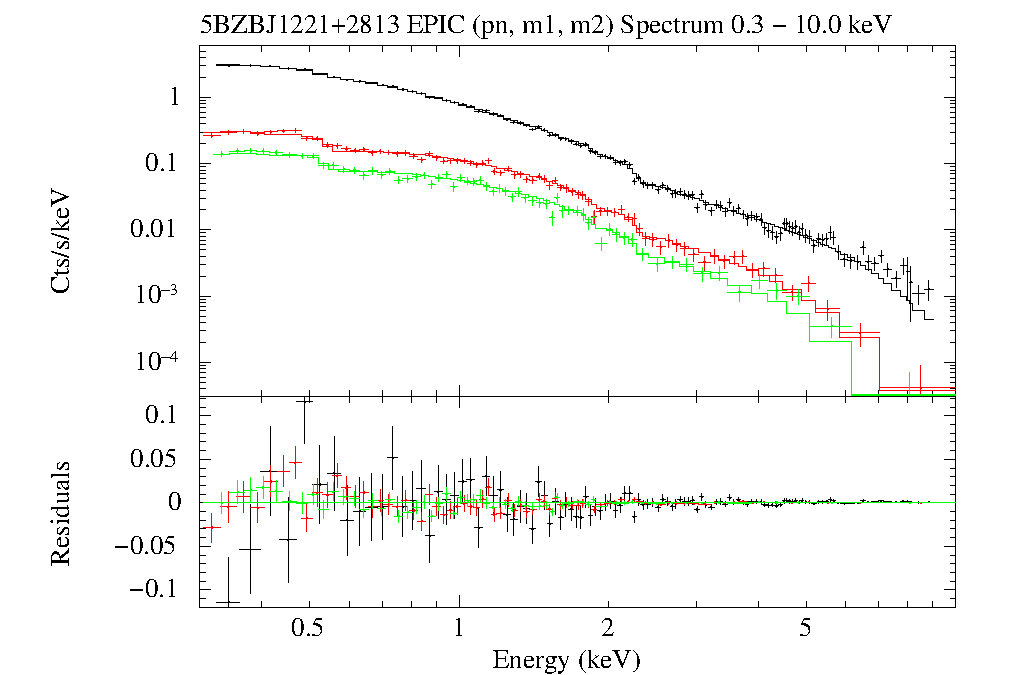}
     \caption{Example light curve and spectra of 5BZBJ1221+2813. {\it Left:} 
     EPIC-pn light curve in the 0.2 - 2.0 keV (top) and 2.0 - 10.0 keV (middle) bands and hardness ratio (bottom). 
     Red symbols indicate those bins for which the effective exposure is below 15\%.
     {\it Right:} Combined spectral fit for EPIC-pn (black), EPIC-MOS1 (green), and EPIC-MOS2 (red)
     spectra corresponding to the {\it power law} with the two-absorption-component model.}
  \label{fig:spectralc}
\end{figure*}

 The SAS task {\tt epiclccorr} is used to produce background-subtracted
source light curves corrected for inefficiencies of the instrument
(vignetting, chip gaps, PSF etc.) and time corrections (dead time, GTIs, etc.). A
hardness ratio (E$_{hard}$/E$_{soft}$) light curve is also produced. The left
panel in Fig.~\ref{fig:spectralc} shows a representative example of a
generated light curve. Weighted average rates over the entire exposure are
derived for the two energy bands and marked in the figure with a discontinuous
line. Time bins below 15\% effective exposure are identified (marked in red)
and not used in the calculation of these average rates.

\subsubsection{X-ray variability}
\label{sec:xrayvar}

\par We measure and quantify the source X-ray variability using the above-generated light curves for each exposure separately and over the two defined energy
ranges ($E_{soft} =$ 0.2 - 2 keV and $E_{hard} =$ 2 - 10 keV). For
both, we use the SAS task {\tt ekstest}. The task uses the source and
background time series to run several variability tests. Only bins within GTI
intervals are used and bins containing null or negative values are eliminated.

\par The $\chi^2$ analysis is used to test the null hypothesis of a constant
source rate, so that sources are reported as variable in
Table C.1 when the $\chi^2$ probability
is $\leq 10^{-5}$  in at least one exposure \citep{Rosen2016}. Both quantities, $\chi^2$ and $\chi^2$
probability, are included in the catalogue for every EPIC exposure and each
energy range ($E_{soft}$ and $E_{hard}$).

\par In addition, to quantify the scale of the variability, we use the
fractional root mean squared variability amplitude $F_{var}$
\citep{1997ApJ...476...70N,2003MNRAS.345.1271V}. This variable is the square
root of the excess variance after normalisation by the mean count rate
$\bar{x}$. This is used as an estimator of the intrinsic source variance and
is defined as:
 \begin{equation}
    F_{var} =\sqrt{\frac{S^2 - \overline{\sigma_{err}^{2}}} {\bar{x}^2} }
 ,\end{equation}

\par
\noindent where $S^2$ is the variance of the time series with N bins,
 \begin{equation}
  S^2 = \frac{1}{N-1} \sum_{i=1}^{N} \left(x_i - \bar{x} \right) ^2
 ,\end{equation}

\par
\noindent and $\overline{\sigma_{err}^2}$ is the mean square error of the time series. $F_{var}$ 
can be expressed as a percentage. The uncertainty in $F_{var}$ takes the form:

 \begin{equation}
   \Delta F_{var}=\frac{1}{2 F_{var}} \sqrt{ \left( \sqrt{\frac{2}{N} }\frac{\overline{\sigma_{err}^2}}{\bar{x}^2} \right)^2 + \left( \sqrt{\frac{\overline{\sigma_{err}^2}}{N}} \frac{2 F_{var}}{\bar{x}}\right)^2} 
 .\end{equation}
 
\par
Both quantities $F_{var}$ and $\Delta F_{var}$ are included in the catalogue for every 
EPIC exposure and each energy range ($E_{soft}$ and $E_{hard}$).

\subsubsection{Spectra generation}\label{src_products_spectra}

\par Source and background spectra are produced in the full energy range (0.2
- 10~keV) with an energy resolution of 5~eV. The SAS task {\tt backscale} is
used to calculate the area of the two regions used to extract the spectra. The
corresponding response and ancillary files are also created. Finally, the spectra
are re-binned in order to avoid over sampling the intrinsic energy resolution of
the EPIC cameras by a factor larger than three while making sure that each
spectral channel contains at least 25 background-subtracted counts to ensure
the validity of Gaussian statistics and hence the applicability of the
$\chi^2$ goodness-of-fit test.

\subsubsection{Spectral fits}
\label{sec:src_products_spectra_fits}

\par All the spectral fits are performed uniformly over our data sample with
the XSPEC package version 12.10.1 \citep{1996ASPC..101...17A}. Spectral
fits to the time-averaged spectra are performed in the energy range 0.3 -10~keV in the case of IMAGING mode exposures and 0.6 - 10~keV in the case of
TIMING mode exposures. Upper and lower confidence intervals are derived for
relevant model parameters at the 68\% level.  For those fits where
$\chi^2_{Norm} > 2.0$, confidence intervals are not derived and instead
parameter errors are reported at the $1\sigma$ level.

\par {\bf For all those instruments available within an observation where the
  source has been detected and passes the quality control flags
  (Section~\ref{quality}), we perform spectral fits and derive spectral
  parameters on an instrument-by-instrument basis and for the combined spectra
  of all those instruments.} 

\par Within XSPEC, fits are performed using C-statistics, while to test the
goodness of fit, that is, how well the statistical model matches the data,
$\chi^2$ statistics is used with standard weights (the statistical error given
in the input spectrum).  All spectra are fitted using the following two
baseline models \citep{2005ApJ...625..727P}:

\begin{itemize}
\item[$\bullet$]{{\it Power law}: 
\begin{equation}
N(E) = e^{-\sigma N_{\rm H}} \cdot K \cdot E^{-\Gamma} \label{baseline1}
,\end{equation}

\par 
\noindent where $\Gamma$ is the photon index.  Within XSPEC, we use the {\tt
  pegpwrlw} model, which also allows  a simple extraction of the flux
at a given energy value. In our case, we extract the X-ray flux at 5~keV.  } \\

\item[$\bullet$]{{\it Log parabola}:
\begin{equation}
N(E) =e^{-\sigma N_{\rm H}} \cdot K \cdot (E/E_1)^{(-\Gamma-\beta \cdot log(E/E_1))} \label{baseline2}
,\end{equation}

\par
\noindent where $\Gamma$ is the photon index and $\beta$ measures the
curvature of the parabola. $E_1$ is a scale parameter (always frozen during
the fitting procedure) which is set to the lower energy range used for fitting
\citep{Massaro2004}. Within XSPEC, the {\tt logpar} model is used.}

\end{itemize}

\par Both models are applied using neutral cold absorption (e$^{-\sigma N_{\rm
    H}}$) where $N_{\rm H}$ is the column density and $\sigma$ is the
photoelectric cross-section of the process according to
\cite{Morrison1983}. For $N_{\rm H}$, the following variations are tested:

\begin{itemize}

\item{$N_{\rm H}$ fixed to the Galactic column density, $N_{\rm H,Gal}$,
  during the fitting procedure. $N_{\rm H,Gal}$ has been taken from the HI4PI,
  a full-sky HI survey based on EBHIS and GASS
  \cite{2016A&A...594A.116H}}. Within XSPEC, the {\tt tbabs} model is used.

\item{$N_{\rm H}$ is left free to vary during the fitting procedure. Within
  XSPEC, the {\tt tbabs} model is used.}

\item{A combination of two $N_{\rm H}$ values is used, $N_{\rm H,Gal}$ and
  $N_{\rm H,Z}$. The first fixed to the Galactic column density value during
  the fitting procedure and the second one used to account for any extra local
  absorption and left free to vary during the fitting procedure. This extra
  local component is a function of the source redshift. Within XSPEC, the {\tt
    ztbabs} model is used.}
  
 \end{itemize}

\par Absorbed fluxes and luminosities (when redshifts are available) are
derived inside XSPEC over the three bands used ({\it soft}, 0.2 - 2.0~keV;
{\it hard}, 2.0 - 10.0~keV; and full energy band 0.2 - 10.0~keV).
The right panel of Fig.~\ref{fig:spectralc} shows an example generated
spectrum and the results of the combined spectral fit using the model {\it  power law} with two absorption components, $N_{\rm H,Gal}$ and $N_{\rm
  H,Z}$.

\subsubsection{Flux upper limits}\label{sec:src_products_ul}

\par When no X-ray source is detected, flux upper limits at the BZCAT source
location are derived over three different energy ranges ({\it soft}, 0.2 -
2.0~keV; {\it hard}, 2.0 - 10.0~keV; and full energy band 0.2 - 10.0~keV) at
the 3$\sigma$ confidence level. Flux upper limits are first derived in cts/sec
from a 30{\arcsec} radius region around the source location using the SAS task
{\tt eregionanalyse} using both a background and an exposure map. Count rates
are converted afterwards to flux in units of 10$^{-11}$~erg~cm$^{-2}$~s$^{-1}$
using Energy Conversion Factors (ECFs) for each energy range
\citep{Mateos2009}. These ECFs are derived from simulations assuming an
absorbed {\it power law} model with $N_H =$ 3.0
10$^{20}$~hydrogen~atoms~cm$^{-2}$ and $Photon Index =$ 1.7. The ECFs used are
camera-, energy band-, and filter dependent. When the redshift is available, upper limits to the luminosity are also derived.  Only two sources in the catalogue do not show significant X-ray emission, namely 5BZBJ1721+6004 and
5BZBJ2036+6553. The source 5BZBJ2214+0020 is present in two {\it XMM--Newton}
observations and in one of them (ObsId 0673000137) no significant emission is
detected. Table~\ref{table:upperlims} summarises these results.

\begin{table*}[h!]
\centering
\caption{List of {\it XMM--Newton} observations where no X-ray emission is detected at the location of the BZCAT source.}
\label{table:upperlims}
\begin{tabular}{lccccccc} 
\hline\hline             
  BZCAT Object 
         & \multicolumn{1}{c}{Observation ID} 
         & \multicolumn{3}{c}{Flux U.L.} 
         & \multicolumn{3}{c}{Luminosity U.L.} \\
  
         & \multicolumn{1}{c}{} 
         & \multicolumn{3}{c}{10$^{-13}~erg~cm^{-2}~s^{-1}$} 
         & \multicolumn{3}{c}{10$^{44}~erg~s^{-1}$} \\
         
         &
         & (0.2-10 keV)
         & (0.2-2 keV)
         & (2-10 keV)
         & (0.2-10 keV)
         & (0.2-2 keV)
         & (2-10 keV)\\
\hline
5BZBJ1721+6004  &  0651371901 & 0.98 & 0.31 & 0.85  & 1.11 & 0.35 & 0.96 \\
5BZBJ2036+6553  &  0301651401 & 7.33 & 2.69 & 10.49 & -- & -- & -- \\
5BZBJ2214+0020(*)  &  0673000137 & 1.60 & 0.70 & 2.13  & -- & -- & -- \\
 \hline
\end{tabular}\\
Column description. (1): Source name as given in BZCAT, (2): {\it XMM--Newton}
Observation ID, (3): flux upper limit at the 3$\sigma$ confidence level in the
three energy ranges indicated, (4): luminosity upper limits {\it idem}. (*)
This BZCAT source is detected in {\it XMM--Newton} Observation ID 0673000136.
\end{table*}



\section{The catalogue}\label{sec:the_catalogue}

\subsection{Description and access}
\label{subsec:the_catalogue_description}

\par The full catalogue is divided into four different tables. Tables B.1,
B.2, B.3, and B.4 report the description of the contents of each one, including
the column name, units, and a short description of each column.

\par The first table is named {\it "catalogue"}, and contains 310 rows, one for
each unique {\it XMM--Newton} observation, each characterised by an observation ID
(ObsId) and 148 columns (Table B.1). Here, for each
individual EPIC camera within an observation, we report  different exposure and analysis
properties, such as analysis quality flags (section~\ref{quality}), source
position rectification (section~\ref{sec:src_posrec}), detection flags (section~\ref{subsec:src_detection}), pile-up and pile-up-correction flags
(section~\ref{subsec:pileup}), fit statistics
(section~\ref{sec:src_products_spectra_fits}), and flags to check whether
light curves and spectral products are extracted
(sections~\ref{src_products_lc} and~\ref{src_products_spectra}). The table
also reports some source properties, such as count rates and variability
parameters as defined in section~\ref{sec:xrayvar}, also the flux extracted
at 5~keV, the total flux, and the luminosity in the entire, soft, and hard
energy bands extracted from the combined EPIC fit for the model {\it power
  law} with two absorption components
(section~\ref{sec:src_products_spectra_fits}). Finally, detection
flags and OM magnitudes and fluxes in the available filters are included
(section~\ref{om_products}).

\par The next table of the {\it catalogue of XMM--Newton BL Lacs } is named
     {\it "model"} and reports information on the fits to the individual
     and combined EPIC exposures. The table contains 7440 rows and 62 columns
     as described in Table B.2. Here we report the best-fit
     model parameters obtained from the spectral fits
     (section~\ref{sec:src_products_spectra_fits}). The {\it model} table also includes
     background-subtracted counts (section~\ref{src_products_lc}) and
     pile-up information (section~\ref{subsec:pileup}). For each ObsId,
     there are six rows corresponding to the {\bf results of the fits for the} six different models and for each one
     of the {\bf three} EPIC instruments (a total of 18 rows). The table also includes six rows,
     one for each of the six different models, for the {\bf results of the} combination of spectra
     for all the EPIC instruments for which there is a valid spectra. The combined spectral fit results are reported as "epic" under the {\tt Inst}
     column of the table. In summary, for each unique observation (ObsId) there are 24 rows (6 models for each one of the three EPIC instruments
     plus their combination).

\par The third extension is {\it "auxdata"} and contains 310 rows, one for
each ObsId and 53 columns with auxiliary data. These include cross-matched
source information, product-extraction regions (source and background),
pile-up-correction regions, EPIC light curve start and end time, light-curve
time bin, and OM source quality flags (section~\ref{om_products}). The
description for these columns is reported in Table B.3.

\par The last table from the {\it catalogue of XMM--Newton BL Lacs } is called
     {\it "multifreq"}, contains 92 rows and 45 columns, and is described in
     Table B.4. This table reports the non-biased ObsIds
     that correspond to 79 different BL Lacs selected as reported in
     Section~\ref{subsec:spectralproperties}. This set of observations is used
     to study the average spectral properties of the sample. The spectral     information included in the table comes from the model {\it log parabola}
     with $N_H$ free. This table also combines this information with
     multi-frequency information as explained in
     section~\ref{subsec:multiwavelength}.

\par In Table C.1 we present a "slim" version of the source
catalogue. The {\it slim} version includes one entry per source in the
catalogue (103) and information on some source properties, including BL Lac
name and coordinates taken from BZCAT, redshift, galactic column density from
the HI4PI, a full-sky HI survey based on EBHIS and GASS
\citep{2016A&A...594A.116H}. The cumulative {\it XMM--Newton} exposure time
available per each BL Lac and the number of observations available in the {\it
XMM--Newton} archive (XSA) for each BL Lac is also included. {\bf Those} BL
Lacs that are not the target of an {\it XMM--Newton} observation but rather
found in the FOV are indicated by a symbol. We also include a flag
indicating whether the light curve and spectra are produced. Additionally, we
report the SED classification, the logarithm of the synchrotron peak in
observed frame as indicated in Section~\ref{sec:sample_selection}, and a flag
indicating whether the source is variable in the soft and hard bands as
described in Section~\ref{sec:xrayvar}. Finally, some multi-wavelength information
is included, such as the name of the 4FGL catalogue association and a flag
indicating whether there is an association in the TeVCat.

\par The full catalogue is provided in a {\bf series of Flexible Image Transport System
(FITS) files with extension names}: {\it "catalogue"}, {\it "model"},
{\it "auxdata",} and {\it "multifreq"}, corresponding to the tables
described above. The {\it catalogue of XMM--Newton BL Lacs} can be found at
VizieR\footnote{\url{https://vizier.u-strasbg.fr}} and will
also be available through ESASky\footnote{\url{https://sky.esa.int}}.

\subsection{Multi-wavelength information}
\label{subsec:multiwavelength}

\par As BL Lacs are the most numerous extragalactic $\gamma$-ray emitters,
we have cross-matched our  catalogue of {\it XMM--Newton BL Lacs} with the 4FGL
\citep{2020ApJS..247...33A} to search for $\gamma$-ray counterparts. A BL Lac
from our catalogue is considered to be associated with a 4FGL source when their angular distance is equal to or smaller than the semi-major axis of the
95\% error ellipse for the position of the 4FGL source.

\par About  61\% (63/ 103) of the {\it XMM--Newton} BL Lacs are present
in the 4FGL, so they are strong $\gamma$-ray emitters. The table {\it
  "multifreq"} includes the associated 4FGL source name, the photon flux
between 1 and 100~GeV (Flux\_1000), the energy flux between 100~MeV and
100~GeV (Energy\_Flux1000), the variability index (Variability\_Index\_4FGL),
and the fractional variability (Frac\_Variability\_4FGL).

\par {\bf Moreover, we report the associations with the
  TeVCat\footnote{\url{http://tevcat.uchicago.edu}} version 3.4 updated up to
  March 2021, which is an online catalogue for TeV astronomy. For the
  identification, we used the same angular distance as for the
  4FGL. About 17.5\% (18/103) of the {\it XMM--Newton} BL Lacs have associated
  TeV emission.}


\par Additionally, we report associations with several radio catalogues. The
radio flux density at 1.4~GHz is selected, performing a cross-match with the
latest version of the radio catalogue NRAO VLA Sky Survey \citep[NVSS,
][]{1998AJ....115.1693C}, as well as that at 845~MHz from the latest release of the Sydney
University Molonglo Sky Survey Source catalogue \citep[SUMSS,
][]{2003MNRAS.342.1117M}. All the cross-matches are performed establishing a
search radius of 10" \citep{2005A&A...430..927G}.

\subsection{Calibration sources}
\label{subsec:ecal_sources}

\par Several of the sources included in our catalogue are ---or were--- part of
the {\it XMM--Newton Routine Calibration Plan},  updated most recently in March 2020
\citep{Smith2020}. These sources have been routinely monitored for instrument
calibration purposes and include: MRK~421, H1426+428, PG1553+11, and
PKS2155-304 (5BZBJ1104+3812, 5BZBJ1428+4240, 5BZBJ1555+1111 and 5BZBJ2158-3013
respectively). Other sources, XMMJ03114-7701, MS06079+7108, MS07379+7441, MS12292+6430, and SDSSJ12350+6205 (5BZBJ0311-7701, 5BZBJ0613+7107,
5BZBJ0744+7433, 5BZBJ1231+6414 and 5BZBJ1237+6258 respectively) have been used
sporadically for calibration purposes over the course of the mission. MRK~421
and PKS2155-304 are used for monitoring the effective area of the RGS
instruments. During these RGS calibration observations, the EPIC instrumental
setup is not optimised for scientific return, which means that in many cases
the reported observations could suffer from irrecoverable pile-up or other
effects.  H1426+428, PG1553+11, and PKS2155-304 are also used for
cross-calibration purposes either between different {\it XMM--Newton}
instruments or with other observatories. In these cases, the EPIC instrumental
setup is optimised for maximum scientific return and the reported observations
in the catalogue can be safely used for science.  Table~\ref{table:cal_obs}
contains the list of calibration observations included
in the catalogue. Here, {\tt RA Nom.}  and {\tt DEC Nom.} correspond to the
nominal pointing of the {\it XMM--Newton} observation rather than the
coordinates of the source.

\begin{table*}[hbt!]
\centering
\tiny
\caption{List of {\it XMM--Newton} calibration observations included in the catalogue.}
\label{table:cal_obs}
\begin{tabular}{lcccrcr} 
\hline\hline             
   \multicolumn{1}{l}{BZCAT Object} 
 & \multicolumn{1}{c}{Observation ID} 
 & \multicolumn{1}{c}{RA Nom.} 
 & \multicolumn{1}{c}{Dec Nom.}   
 & \multicolumn{1}{r}{Revolution} 
 & \multicolumn{1}{c}{Obs. Start} 
 & \multicolumn{1}{r}{Obs. Duration} \\ 
 &                
 & \multicolumn{1}{c}{HH MM SS.SS} 
 & \multicolumn{1}{c}{DD MM SS.S} 
 & 
 & \multicolumn{1}{c}{YYYY-MM-DD HH:MM:SS.S} 
 & \multicolumn{1}{c}{secs} \\ 
\hline
5BZBJ0311-7701& 0122520201& 03 11 54.99& -76 51 52.0&     57& 2000-03-31 10:56:11.0&   45920\\
5BZBJ0613+7107& 0656580301& 06 15 36.29& +71 02 15.0&   2246& 2012-03-15 11:27:22.0&   35696\\
5BZBJ0744+7433& 0123100101& 07 44 04.49& +74 33 49.5&     63& 2000-04-13 03:42:40.0&   66662\\
              & 0123100201& 07 44 04.49& +74 33 49.5&     63& 2000-04-12 19:02:56.0&   23600\\
5BZBJ1104+3812  &       0153951201& 11 04 23.01& +38 10 43.0&   1083& 2005-11-07 19:57:22.0&   10017\\
              & 0791781401& 11 04 24.99& +38 12 44.0&   3287& 2017-11-19 09:27:28.0&   69800\\
              & 0791782001& 11 04 24.99& +38 12 44.0&   3373& 2018-05-09 21:33:12.0&   72960\\
              & 0810860201& 11 04 24.99& +38 12 44.0&   3471& 2018-11-21 12:34:13.0&   70300\\
              & 0810860701& 11 04 24.99& +38 12 44.0&   3554& 2019-05-06 02:44:58.0&   64800\\
              & 0810862501& 11 04 24.99& +38 12 44.0&   3658& 2019-11-29 07:09:40.0&   61000\\
              & 0810863001& 11 04 24.99& +38 12 44.0&   3735& 2020-04-30 20:51:11.0&   69500\\
              & 0411081601& 11 04 26.61& +38 12 37.5&   1358& 2007-05-10 15:44:21.0&    8908\\
              & 0158970101& 11 04 27.29& +38 12 31.8&    637& 2003-06-01 11:33:26.0&   47538\\
              & 0158970201& 11 04 27.29& +38 12 31.8&    637& 2003-06-02 01:03:59.0&    8963\\
              & 0158971201& 11 04 27.29& +38 12 31.8&    807& 2004-05-06 02:38:06.0&   66141\\
              & 0158971301& 11 04 27.29& +38 12 31.8&   1084& 2005-11-09 18:18:04.0&   60015\\
              & 0162960101& 11 04 27.29& +38 12 31.8&    733& 2003-12-10 21:23:14.0&   50755\\
              & 0411080301& 11 04 27.29& +38 12 31.8&   1184& 2006-05-28 02:07:55.0&   69212\\
              & 0411080701& 11 04 27.29& +38 12 31.8&   1280& 2006-12-05 11:46:57.0&   18907\\
              & 0411081301& 11 04 27.29& +38 12 31.8&   1358& 2007-05-10 03:37:41.0&   18913\\
              & 0411081901& 11 04 27.29& +38 12 31.8&   1455& 2007-11-19 13:09:14.0&   18910\\
              & 0411082701& 11 04 27.29& +38 12 31.8&   1552& 2008-05-31 02:07:53.0&   71475\\
              & 0411083201& 11 04 27.29& +38 12 31.8&   1820& 2009-11-16 17:37:59.0&   58070\\
              & 0510610101& 11 04 27.29& +38 12 31.8&   1357& 2007-05-08 14:42:08.0&   27654\\
              & 0510610201& 11 04 27.29& +38 12 31.8&   1357& 2007-05-08 08:05:31.0&   22706\\
              & 0560980101& 11 04 27.29& +38 12 31.8&   1640& 2008-11-22 14:07:29.0&   71318\\
              & 0560983301& 11 04 27.29& +38 12 31.8&   1732& 2009-05-25 03:37:32.0&   64173\\
              & 0656380101& 11 04 27.29& +38 12 31.8&   1904& 2010-05-03 07:19:29.0&   51169\\
              & 0656380801& 11 04 27.29& +38 12 31.8&   2001& 2010-11-12 20:51:05.0&   42669\\
              & 0656381301& 11 04 27.29& +38 12 31.8&   2002& 2010-11-14 20:44:26.0&   42521\\
              & 0658800101& 11 04 27.29& +38 12 31.8&   2094& 2011-05-17 10:10:06.0&   35074\\
              & 0658800801& 11 04 27.29& +38 12 31.8&   2192& 2011-11-28 23:28:36.0&   26673\\
              & 0791780101& 11 04 27.29& +38 12 32.0&   3096& 2016-11-03 13:15:45.0&   17500\\
              & 0791780601& 11 04 27.29& +38 12 32.0&   3187& 2017-05-04 04:01:33.0&   12500\\
              & 0158970701& 11 04 27.30& +38 12 31.8&    640& 2003-06-07 20:44:27.0&   10900\\
              & 0658801301& 11 04 27.30& +38 12 31.8&   2837& 2015-06-05 23:48:35.0&   29000\\
              & 0658801801& 11 04 27.30& +38 12 31.8&   2915& 2015-11-08 13:42:37.0&   33600\\
              & 0658802301& 11 04 27.30& +38 12 31.8&   3005& 2016-05-06 03:38:20.0&   29400\\
              & 0411081501& 11 04 28.36& +38 12 23.2&   1358& 2007-05-10 12:37:41.0&    8909\\
              & 0411081401& 11 04 29.39& +38 12 14.6&   1358& 2007-05-10 09:31:01.0&    8910\\
              & 0153951301& 11 04 31.58& +38 14 20.6&   1083& 2005-11-07 16:37:22.0&    9719\\
              & 0153950601& 11 04 36.40& +38 11 16.0&    440& 2002-05-04 16:09:17.0&   39727\\
              & 0153950701& 11 04 42.29& +38 12 55.0&    440& 2002-05-05 03:51:30.0&   19982\\
5BZBJ1231+6414& 0124900101& 12 31 32.01& +64 14 21.0&     82& 2000-05-21 01:10:33.0&   71450\\
              & 0158560301& 12 31 32.01& +64 14 21.0&    621& 2003-05-01 08:24:13.0&   60641\\
5BZBJ1428+4240& 0165770101& 14 28 32.66& +42 40 20.6&    852& 2004-08-04 00:59:26.0&   67866\\
              & 0165770201& 14 28 32.66& +42 40 20.6&    853& 2004-08-06 00:32:43.0&   68920\\
              & 0212090201& 14 28 32.66& +42 40 20.6&    939& 2005-01-24 14:44:40.0&   30417\\
              & 0310190101& 14 28 32.66& +42 40 20.6&   1012& 2005-06-19 07:39:40.0&   47034\\
              & 0310190201& 14 28 32.66& +42 40 20.6&   1015& 2005-06-25 06:03:28.0&   49505\\
              & 0310190501& 14 28 32.66& +42 40 20.6&   1035& 2005-08-04 04:52:10.0&   47542\\
5BZBJ1555+1111& 0727780101& 15 55 42.99& +11 11 24.4&   2495& 2013-07-24 14:57:49.0&   34500\\
              & 0727780201& 15 55 42.99& +11 11 24.4&   2680& 2014-07-28 04:00:06.0&   36300\\
              & 0810830101& 15 55 42.99& +11 11 24.4&   3427& 2018-08-25 15:35:27.0&   35100\\
              & 0810830201& 15 55 42.99& +11 11 24.4&   3611& 2019-08-27 15:52:27.0&   32950\\
              & 0727780301& 15 55 43.03& +11 11 24.3&   2882& 2015-09-04 18:23:24.0&   29999\\
              & 0727780401& 15 55 43.03& +11 11 24.3&   3057& 2016-08-17 21:56:06.0&   30000\\
              & 0727780501& 15 55 43.03& +11 11 24.3&   3242& 2017-08-22 15:00:00.0&   30020\\
\hline
\end{tabular} 
\end{table*}

\begin{table*}[hbt]
\centering
\tiny
\caption*{{\it Continuation}, list of {\it XMM--Newton} calibration observations included in the catalogue.}
\begin{tabular}{lcccrcr} 
\hline\hline             
   \multicolumn{1}{l}{BZCAT Object} 
 & \multicolumn{1}{c}{Observation ID} 
 & \multicolumn{1}{c}{RA Nom.} 
 & \multicolumn{1}{c}{Dec Nom.}   
 & \multicolumn{1}{r}{Revolution} 
 & \multicolumn{1}{c}{Obs. Start} 
 & \multicolumn{1}{r}{Obs. Duration} \\ 
 &                
 & \multicolumn{1}{c}{HH MM SS.SS} 
 & \multicolumn{1}{c}{DD MM SS.S} 
 & 
 & \multicolumn{1}{c}{YYYY-MM-DD HH:MM:SS.S} 
 & \multicolumn{1}{c}{secs} \\ 
\hline    
5BZBJ2158-3013& 0124930501& 21 58 52.06& -30 13 32.1&    450& 2002-05-24 09:31:02.0&  104868\\
              & 0124930601& 21 58 52.06& -30 13 32.0&    545& 2002-11-29 23:27:28.0&  114675\\
              & 0158960101& 21 58 52.06& -30 13 32.1&    724& 2003-11-23 00:46:22.0&   27159\\
              & 0158960901& 21 58 52.06& -30 13 32.1&    908& 2004-11-22 21:35:30.0&   28919\\
              & 0158961001& 21 58 52.06& -30 13 32.1&    908& 2004-11-23 19:45:55.0&   40419\\
              & 0158961101& 21 58 52.06& -30 13 32.1&    993& 2005-05-12 12:51:06.0&   28910\\
              & 0158961301& 21 58 52.06& -30 13 32.1&   1095& 2005-11-30 20:34:03.0&   60415\\
              & 0158961401& 21 58 52.06& -30 13 32.1&   1171& 2006-05-01 12:25:55.0&   64814\\
              & 0411780101& 21 58 52.06& -30 13 32.1&   1266& 2006-11-07 00:22:47.0&  101012\\
              & 0411780201& 21 58 52.06& -30 13 32.1&   1349& 2007-04-22 04:07:23.0&   67911\\
              & 0411780301& 21 58 52.06& -30 13 32.1&   1543& 2008-05-12 15:02:34.0&   61216\\
              & 0411780401& 21 58 52.06& -30 13 32.1&   1734& 2009-05-28 08:08:42.0&   64820\\
              & 0411780501& 21 58 52.06& -30 13 32.1&   1902& 2010-04-28 23:47:42.0&   74298\\
              & 0411780601& 21 58 52.06& -30 13 32.1&   2084& 2011-04-26 13:50:40.0&   63818\\
              & 0411780701& 21 58 52.06& -30 13 32.1&   2268& 2012-04-28 00:48:26.0&   68735\\
              & 0411782101& 21 58 52.06& -30 13 32.1&   2449& 2013-04-23 22:31:38.0&   76015\\
              & 0727770901& 21 58 52.09& -30 13 32.1&   2633& 2014-04-25 03:14:56.0&   65000\\
              & 0124930101& 21 58 53.00& -30 13 35.0&     87& 2000-05-30 05:29:42.0&   61109\\
              & 0124930201& 21 58 53.00& -30 13 35.0&     87& 2000-05-31 00:30:51.0&   72558\\
              & 0124930301& 21 58 53.00& -30 13 35.0&    362& 2001-11-30 02:36:09.0&   92617\\
\hline
\end{tabular}\\
Column description. (1): Source name as given in  BZCAT, (2): {\it XMM--Newton} observation ID, (3): right ascension of nominal {\it XMM--Newton} pointing, (4): declination of nominal {\it XMM--Newton} pointing, (5): revolution number, (6) observation start date and (7): observation duration. 
\end{table*}

\subsection{Sources with long cumulative exposure times}
\par There are 19 sources in our catalogue with cumulative exposure times
longer than 100~ks as a result of the existence of several observations spread out in time. Some of these sources are of potential interest
for  studies of long-term variability. Five of these sources are the above-mentioned calibration sources MRK~421, PG1553+11, PKS2155-304, H1426+428, and
MS12292+6430 (5BZBJ1104+3812, 5BZBJ1555+1111, 5BZBJ2158-3013, 5BZBJ1428+4240
and 5BZBJ1231+6414, respectively).

\par Another six correspond to some well-studied BL Lacs, such as
5BZBJ0854+2006 (PKS 0537-441, 364.6 ks) which is a variable BL Lac
\citep{2007ApJ...664..106P}. 5BZBJ1031+5053 (1ES 1028+511, 307.3 ks) has
been observed by Fermi and has been studied at several wavelengths
\citep{2007A&A...466...63W,2009MNRAS.398..832K}. 5BZBJ1221+2813 (W Comae,111.6 ks) has been detected in TeV
\citep{2008ApJ...684L..73A}. 5BZBJ1653+3945 (Mrk501, 1946.7 ks) is a {\bf very high-energy (VHE)} BL
Lac \cite{2020ApJ...901..132S,2021MNRAS.504..878D}. 5BZBJ2202+4216 (195.4ks)
is the famous {\bf BL Lacertae}, the prototype of the BL Lac class AGN
\citep{1968Natur.218..663S,1969Natur.223..566O,1971PASP...83..680P}. 5BZBJ2359-3037
(H2356-309, 687.6 ks) has been followed up because of its VHE TeV emission
\citep{2006A&A...455..461A,2010A&A...516A..56H}.

\par The remaining eight sources are serendipitous BL Lacs that happen to be
in the FOV of {\it XMM--Newton} observations. This is the case for
5BZBJ0057-2212 (219.7 ks), 5BZBJ0333-3619 (1094.4 ks), 5BZBJ0613+7107
(267.2 ks; additionally, this BL Lac has been used sporadically for
calibration purposes), 5BZBJ0721+7120 (172.4 ks; in some observations this
source is the target of {\it XMM--Newton} observations), 5BZBJ0958+6533 (159.6
ks), 5BZBJ1136+1601 (140.7 ks), 5BZBJ1210+3929 (167.8 ks), and
5BZBJ2258-3644 (170.3 ks). Appendix A collects some basic information on
these sources.

\section{Results}\label{sec:results}
\subsection{Redshift distribution}

\par This catalogue presents a uniform analysis of 310 observations covering
nearly 20 years of {\it XMM--Newton} archival public observation of 103 BL Lacs
up to September 2019. From these observations, we are able to extract and fit
the spectra for 271/307 ($\sim$ 88\%) of them, which are used to derive
spectral properties of the sample. Light curves are extracted for 305/307($\sim$99\%) of the observations, which are used to apply statistical tools to
estimate and quantify the flux variability of the sources in the catalogue.

\par
\begin{figure}[!htb]
  \centering
  \includegraphics[width=.45\textwidth]{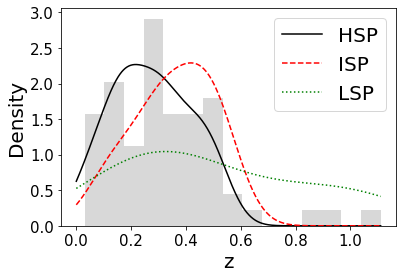}
  \caption{Redshift distribution for the sources in the {\it XMM--Newton BL Lac catalogue}.  Grey bars represent all BL Lacs for 
  which the redshift is known (60\% of the total sample), while curves represent the KDE curves for the subpopulations 
  HSP (solid black), ISP (dashed red), and LSP (dotted green).}
  \label{fig:histredshift}
\end{figure}

 Fig.~\ref{fig:histredshift} shows the distribution of redshift for the
62/103 ($\sim$60\%) BL Lacs for which redshift information is available. The same plot shows the
kernel density estimate (KDE) curves for the distribution of redshifts when
the sources are classified according to HSPs, ISPs, and LSPs. The sample
includes objects with redshift up to 1.11.

\par In order to derive average X-ray spectral parameters from the best-fit
model, we need to establish first which spectral model out of the six tested
best fits the data on average
(Section~\ref{sec:src_products_spectra_fits}). Also, we need to remove any
bias introduced by the existence of many observations from certain sources. In
the coming sections, we discuss the selection of the best-spectral-fit model and
then the average X-ray spectral properties derived from it.

\subsection{Best-fit model}

\par In order to determine which spectral model best fits the data, we use the
fit results from the combination of the EPIC detectors, indicated as "epic" in
column {\it Inst} of Table B.2, which are free or successfully
corrected for pile-up. For every single observation, this can be any
combination of one, two, or three EPIC cameras.  This leaves a total of 218
observations, which are used to extract results for each one of the six
available models.

\par
\begin{figure*}[h!]
  \centering
    \includegraphics[width=.33\textwidth]{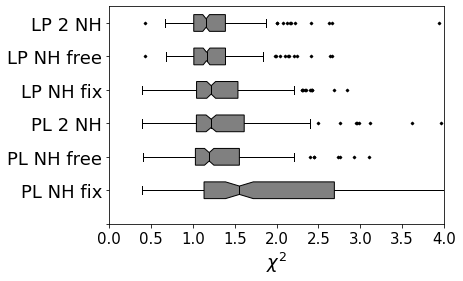}
    \includegraphics[width=.33\textwidth]{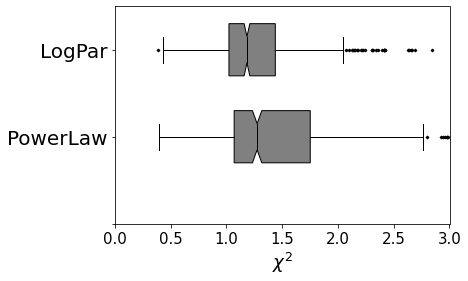}
    \includegraphics[width=.33\textwidth]{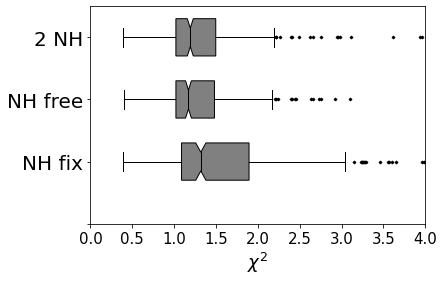}
    \caption{$\chi_r^2$ box plots. In each panel, the median value for each
      model is depicted by a vertical line inside the box, the grey boxes      range from the first quantile (Q1) to the third quantile (Q3), the notch
      around the median indicates the confidence interval (1.5 $\times$ (Q3 -
      Q1)), the lines extending horizontally from the boxes ({\it whiskers})
      indicate the minimum and maximum value after removing outliers. Outliers
      are plotted as stars.  {\it Left:} Six models from top to bottom: {\it
        log parabola} with a combination of $N_{H,Gal}$ and $N_{H,z}$, $N_H$
      free and $N_H$ fixed to $N_{H,Gal}$, {\it power law} with a combination
      of $N_{H,Gal}$ and $N_{H,z}$, $N_H$ free and $N_H$ fixed to
      $N_{H,Gal}$. {\it Middle:} From top to bottom: {\it log parabola} and
      {\it power law} models. {\it Right:} From top to bottom: models with
      combination of $N_{H,Gal}$ and $N_{H,z}$ $, N_H$ free, and $N_H$ fixed to
      $N_{H,Gal}$.}
  \label{fig:boxplots}
\end{figure*}

 The {\it power law} and {\it log parabola} models with the different flavours of N$_H$ are fitted uniformly to the entire sample. The goodness of
fit is evaluated using $\chi^2$ statistics. Figure~\ref{fig:boxplots} presents a
box plot of the distribution of $\chi_r^2$ for the six models applied and
described in Section~\ref{sec:src_products_spectra_fits}. This type of plot is
a non-parametric method for graphically representing variation in samples of a
statistical population without making any assumptions as to the underlying
statistical distribution. These plots show the median, the upper (Q3, 75\%),
and lower (Q1, 25\%) quantiles, the minimum and maximum values after excluding
possible outliers (indicated in the figure by the whiskers), the confidence
interval (represented as 1.5 $\times$ (Q3 - Q1) and indicated in the figure by
a notch), and outliers (indicated in the figure by the stars).

\par Figure~\ref{fig:boxplots} shows that the best models in terms of goodness
of fit are both the {\it log parabola} with N$_H$ free and the {\it log
  parabola} with two $N_H$ components ($N_{H,Gal}$ and $N_{H,z}$). Of the
six models tested, the {\it log parabola} with N$_H$ free provides the best
fit in 71/218 cases, followed by the {\it log parabola} with two $N_H$
components (57/218). The {\it power law} model with two absorption components
follows (35/218) and then the {\it log parabola} fixing $N_H$ (24/218). The
{\it power law} with $N_H$ free (19/218) and fixing $N_H$ (12/218) provide the
least number of times the best representation of the
data. Figure~\ref{fig:boxplots} also shows that, on average, {\it log parabola}
models provide a better fit to the data than {\it power law} models, with lower medians and less dispersion of $\chi_r^2$. A {\it log parabola} is
favoured in 152/218 of the spectra fitted versus a simple {\it power law }which is favoured in 66/218. In terms of $N_H$, using two $N_H$ components
($N_{H,Gal}$ and $N_{H,z}$) provides better fits (92/218), while leaving the
$N_H$ free to vary during the fitting procedure is favoured in 90/218 spectra
and fixing $N_H$ to the $N_{H,Gal}$  is favoured in 36/218 spectra.

%

\par
\begin{figure}[h!]
  \centering
  \includegraphics[width=.45\textwidth]{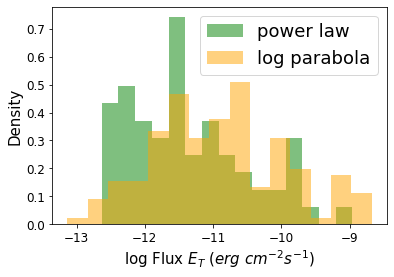}
  \caption{Flux distribution in the band 0.2 - 10.0 keV (in green) for the
    sources best fitted by the {\it power law} model, and the same for the
    {\it log parabola} {\bf (in orange) combining all three absorption flavours.}}
  \label{fig:fluxmodels}
\end{figure}

 In Fig.~\ref{fig:fluxmodels} we represent the flux distributions in the
0.2 - 10.0 keV band for those sources for which the {\it power law} model is preferred and
those best fitted by a {\it log parabola} (combining all three absorption flavours), and see that those sources favouring a {\it log parabola} present
significantly higher fluxes. We use a t-test and reject the null hypothesis
that the means of both distributions are equal considering different variances
with 99\% confidence. {\bf The fact that the {\it log parabola} is preferred for
sources that show higher fluxes, that is, sources with higher S/N as in
\cite{2005A&A...433.1163D}, might indicate that curvature is intrinsic
to this type of object but only shows when the S/N is sufficiently high.}

%
%

\subsection{Spectral properties}
\label{subsec:spectralproperties}
\par For the following analysis, we only consider the best-fit model {\it log
  parabola} with $N_H$ free, and we exclude from this sample those ObsIds poorly fitted with a value for the $\chi_r^2 \geq$ 2 and those ObsIds that
show pile-up in any of the EPIC cameras (see
Table~\ref{table:pileup_exp}). After filtering, we have 200 ObsIds that
correspond to 79 different BL Lacs. To avoid over-representation by sources
observed multiple times we adopt the following procedure
\citep{2005A&A...433.1163D}:

\begin{itemize}
    \item For non-variable sources that only show small variations in both
      photon index and flux in the 0.2 - 10 keV band, that is, $\Delta \Gamma
      \leq$ 0.5 and flux variation of a factor of less than two between maximum and
      minimum, we use the mean value for all output parameters derived using
      the {\it log parabola} with $N_H$ free model.  Given that XSPEC returns
      asymmetric errors, the mean and its error are calculated as follows: for
      each of the 200 ObsIds, we create a Gaussian distribution centred at the
      parameter's value and a total dispersion with an amplitude of the asymmetric
      error, and take a value inside this distribution. We repeat this process
      1000 times and choose as the mean parameter the mean of all
      simulations for each BL Lac.

    \item For variable sources that show significant photon index and/or flux
      variations, that is, $\Delta \Gamma >$ 0.5 and/or flux variation of a factor
      of greater than two between maximum and minimum, we consider the observations
      corresponding to the extreme parameters, so that the source is represented in
      both states.
\end{itemize}

\par The result of this exercise is what we refer to as a non-biased sample and
is summarised in the fourth table of the catalogue, named {\it "multifreq"} as
explained in Section~\ref{subsec:the_catalogue_description}, and its columns
are reported and described in Table B.4.


\par
\begin{figure}[h]
  \centering
    \includegraphics[width=.45\textwidth]{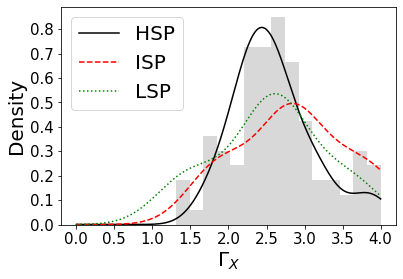}
    \includegraphics[width=.45\textwidth]{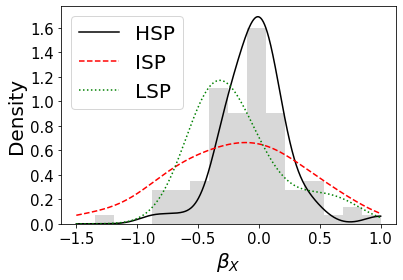}
     \caption{{\it Top:} {\it Log parabola} N$_H$ free model photon index
       $\Gamma_X$ distribution for the non-biased sample of BL Lacs (grey
       bars), and curves {\bf showing} the KDE for the subpopulations HSP (black solid
       line), ISP (dashed red line), and LSP (green dotted line). {\it
         Bottom:} Idem $\beta_X$ distribution. }
  \label{fig:histlogpar}
\end{figure}

 We study the model parameters separating them according to BL Lac
subpopulations, seeing that the photon index is steepest for ISPs, followed by LSPs,
and lastly HSPs ($\Gamma = $ 2.84, 2.59 and 2.49, respectively, as seen in the
top panel of Fig.~\ref{fig:histlogpar}). The ISP distribution is shifted
towards higher $\Gamma_X$ values, however this effect might be affected by
the source 5BZBJ0021-2550, for which $\Gamma_X =$ 3.99, although the model clearly
does not represent  the data well enough as $\chi_r^2 =$ 0.6124.  As for the
curvature parameter $\beta_X$, LSPs are those showing the lowest values
(distribution peaks at -0.289) while ISPs and HSPs show values for the peak of
the distribution at -0.209 and -0.013 respectively, as seen in the bottom panel of Fig.~\ref{fig:histlogpar}. Only a few sources have positive $\beta_X$
values, which could indicate an upward curvature due to the presence of both
synchrotron and inverse Compton components \citep{2002babs.conf...63G}.

\subsection{Luminosity}
\par
\begin{figure}
  \centering
  \includegraphics[width=.45\textwidth]{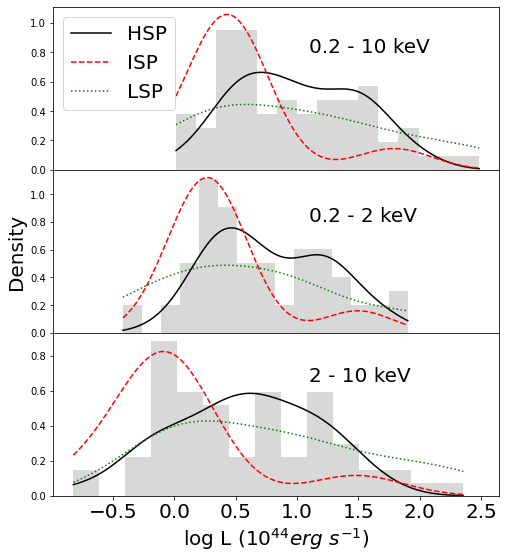}
  \caption{Luminosity distribution of the non-biased sample in the energy
    intervals 0.2 - 10.0 keV (top), 0.2 - 2.0 keV (middle), and 2.0 - 10.0
    (bottom). Grey bars represent sources with available redshift, while
    curves represent the KDE for HSPs (solid black), ISPs (dashed red), and
    LSPs (dotted green).}
  \label{fig:histL}
\end{figure}

 The distribution of the X-ray luminosity in the intervals 0.2 - 10.0 keV,
0.2 - 2.0 keV, and 2.0 - 10.0 keV for the non-biased sample is given in
Fig.~\ref{fig:histL} in the top, middle, and bottom panels, respectively.  The
peak of the distribution for BL Lacs in the 0.2 - 10.0 keV band falls at
$\tilde{L} =$ 8.33$^{+0.07}_{-0.33} \cdot 10^{44}$ erg s$^{-1}$. LSPs are the
most luminous BL Lacs in X-rays, where the median values in the 0.2 - 10.0 keV
band, {\bf are} $\tilde{L}_{LSP} =$ 11.03, $\tilde{L}_{HSP} =$ 9.58, and $\tilde{L}_{ISP}
=$ 2.82 in units of $10^{44} \cdot$ erg s$^{-1}$. This is true for both soft
and hard bands.

\subsection{Variability}

\par The fraction of BL Lacs reported as variable in either band according to
the $\chi^2$ probability of the constancy test is 56/103 ($\sim$ 54\%).  These
sources are mostly HSPs (31), but also there are 9 ISPs and 14 LSPs that are
variable, and 5 of them remain unclassified. The peak of the synchrotron
emission $\nu_{peak}^{S}$ for variable BL Lacs lies at higher values
($log~\tilde{\nu}_{peak \ var}^{S} =$ 15.35) than for non-variable BL Lacs
($log~\tilde{\nu}_{peak \ nonvar}^{S} =$ 14.78). Additionally, most of the
sources that are strong $\gamma$-ray emitters are variable (13/18 TeV sources
and 42/63 4FGL sources).

\section{Conclusions}
\label{sec:conclusions}

\par This paper presents the {\it XMM--Newton BL Lac} catalogue, which comprises a sample of those BL Lacs reported by BZCAT observed with {\it XMM--Newton}
over almost 20 years of the mission. We report results from a total of 310
observations that contain information on 103 different BL Lacs.

\par We have developed an analysis pipeline to automatically cross-match
sources from the BZCAT catalogue and {\it XMM--Newton} public observations contained in the XSA archive as of September 2019. Source-matching and
detection algorithms yield 310 positive X-ray source identifications which
have been homogeneously analysed to generate {\it XMM--Newton} images, spectra,
and light curves from the three EPIC cameras. OM exposures within these
observations are also reduced when available and visual magnitudes and fluxes
are provided.

\par While the sources contained in the {\it XMM--Newton BL Lac} catalogue are
also present in the 4XMM, our pipeline has been specifically designed for the analysis of BL
Lacs. The main difference can be found in the derivation of the
reported fluxes. The pipeline used to extract the 4XMM is optimised for source
detectability and to homogeneously account for all the different classes of
sources included in the catalogue. Unlike in the {\it XMM--Newton BL Lac}
  catalogue, 4XMM sources are not pile-up corrected and the fluxes are derived
from energy conversion factors (ECFs), assuming a spectral model of an
absorbed {\it power law} with absorbing column density $Nh = 3.0 \cdot 10^{20} cm^2$
and continuum spectral slope $\Gamma$ = 1.7 \citep{Rosen2016}.

\par We extracted close to 1000 background-subtracted light curves over
two different energy ranges, soft and hard, together with time variability
information following statistical indicators. Half of our sample shows
indication for variable X-ray emission, most attributed to HSPs.
Regarding the spectra, we performed around 7000 spectral fits to
extract the best-fit model X-ray spectral parameters for two different models, namely
{\it power law} and {\it log parabola,} with three different flavours for the
absorption column density (fixed to the galactic column density, left free to
vary, and a combination of galactic column density plus extra local
absorption). We observe a wide span of photon indices, from the very
flat hard values indicating inverse-Compton-dominated spectra up to very steep
soft indices where the tail of the synchrotron emission reaches the
X-rays. The curvature parameter is mostly negative and lower for sources with
lower $\nu_{peak}$, reaching minimum values for LSPs.

\par
As in previous works
\citep{2002babs.conf...63G,2005A&A...433.1163D,2005ApJ...625..727P} we find
that the model that best represents the X-ray spectra of BL Lacs is the {\it
  log parabola} with $N_H$ left free to vary. Moreover, we see that models
best fit by a {\it log parabola} show higher fluxes than those best
represented by a {\it power law}, indicating that curvature is intrinsic to
this type of object; however we are only able to evaluate curvature when spectra have
high S/N. \cite{Massaro2004} explained the {\it log parabolic} spectrum as the
result of a statistical acceleration mechanism in which the probability of a
particle to remain inside an acceleration region decreases as the energy of
the particles increases.

\par The {\it XMM--Newton BL Lac} catalogue is intended to be used as a
valuable resource to better understand BL Lacs in the X-rays and their
correlations with other wavelengths. Incremental releases are planned to
augment this catalogue, with continuous efforts made by the
astronomical community in discovering new BL Lacs and more observations
becoming available in the {\it XMM--Newton} archive in coming years. {\it
  XMM--Newton} instruments remain in good health and are expected to continue
being operational for several more years.


\begin{acknowledgements}
We thank the anonymous referee for her/his insightful comments that led to
significant improvements in the paper.  IC would like to thank the XMM--Newton
SOC and its members for many fruitful discussions. IC would like to
acknowledge support by the Torres Quevedo fellowship programme from the
Ministerio de Educación y Ciencia Español and from INSA, and to N. Loiseau who
proposed this fellowship. NAC is supported by European Space Agency (ESA)Research Fellowships. This work is based on observations with XMM--Newton, an
ESA science mission with instruments and contributions directly funded by ESA
Member States and NASA. This research has made use of data archives,
catalogues and software tools from the ASDC, a facility managed by the Italian
Space Agency (ASI). This research has made use of the NASA/IPAC Extragalactic
Database, operated by the Jet Propulsion Laboratory, Caltech, under the
contract with NASA, and the NASA Astrophysics Data System Abstract Service.
\end{acknowledgements}

\bibliographystyle{aa} 
\bibliography{biblio}


\section*{Appendix A: Sources with long cumulative exposure times.}\label{AppA}
\par Nineteen BL Lacs in our catalogue have long cumulative exposure
times of over 100~ks, as a result of the existence of several observations, in
some cases, spread out over several years. Amongst these BL Lacs, eight are
serendipitous sources contained in the FOV of {\it XMM--Newton}
target observations. Most of these eight sources are contained in different
samples of objects selected according to different properties. Below we
summarise some of their X-ray spectral properties.

\par {\bf 5BZBJ0057-2212} Of unknown redshift, there are four observations
contained in the catalogue (2000, 2002, 2015 and 2016). All observations but
one show a weak X-ray source for which it is not possible to extract good-quality spectra. For the observation with best statistics (2015), the optimal
fit is provided by a {\it power law} model with two absorption components. The photon
index is 1.84 with a flux of $3.24 \cdot 10^{-13}~erg~cm^{-2}~s^{-1}$ in the
0.2 - 10 keV range. The source shows no sign of variability within the timescale of the observations for the default 500 s bins used.

\par {\bf 5BZBJ0333-3619} This BL Lac is observed at $z = $ 0.308. Twelve
observations spanning 2003 - 2013 are included in the catalogue and all but
one are best fitted by a log parabolic model. The average flux is about
$0.5 - 3.5 \cdot 10^{-12}~erg~cm^{-2}~s^{-1}$ in the 0.2 - 10 keV range. This
source is part of a sample of AGNs that have been the target of TeV
observations by H.E.S.S. up to 2011 that show no significant excess in high-energy emission \citep{Abramowski2014}. There is no significant variability
within the timescale of the observations for the default 500 s bins used.

\par {\bf 5BZBJ0613+7107} The redshift of this source is 0.267. Sixteen
observations from 2000 to 2015 are contained in the catalogue. Eight
observations show a weak X-ray source for which it is not possible to extract
good-quality spectra. For the remaining observations, the most favoured fit is
provided by a {\it power law} model with two absorption components, although in
several observations a log parabolic model fits better. The flux ranges
between $0.6 - 1.3 \cdot 10^{-12}~erg~cm^{-2}~s^{-1}$ in the 0.2 - 10 keVrange. There is no significant variability within the timescale of the
observations for the default 500 s bins used.

\par {\bf 5BZBJ0721+7120} (also known as PKS 0716+714) has been extensively studied
because of its intraday variability. {\bf Simultaneous} radio, optical, UV, and X-ray
observations yielded a short duty cycle of variability at all frequencies and
a correlation between different wavelengths \citep{2006A&A...456..117A}. VLBI
images show a compact, one-sided core jet \citep{2006A&A...452...83B}. Of unknown redshift, five observations are found between 2001 and 2007 in our
catalogue. In the 2001 and 2002 observations, this is a field source while in
2004 and 2007 it was the {\it XMM--Newton} target, part of a target of
opportunity (ToO) program to explore blazars in outburst. During 2004 and 2007
the source displayed high flux variability within the {\it XMM--Newton}
observations, with rapid flares of the order of a few hours
\citep{Kushwaha2020}. During these observations the average flux was of the
order of $1.0 \cdot 10^{-11}~erg~cm^{-2}~s^{-1}$ in the 0.2 - 10 keV range. In
all observations the best fit is provided by a log parabolic model.

\par {\bf 5BZBJ0958+6533} This is a LSP BL Lac at $z =$ 0.367 and famous for
its intraday variability \citep{2006A&A...452...83B}. It shows a one-sided jet
structure that bends from milliarcsecond to arcsecond scales
\citep{2016A&A...594A..60L}. There are three observations contained in our
catalogue, one from 2005 and two from 2007. In all cases, the log parabolic
model provides the best fit. The X-ray flux in the 0.2 - 10 keV range varies
between 2.5 and 3.3$ \cdot 10^{-13}~erg~cm^{-2}~s^{-1}$. The first observation
in 2007 shows a significant steady increase, a factor two, in flux over the course of the observation, more noticeable in the low 0.2-2.0~keV energy
range, which results in a softening of the spectrum over the course of the
observation.

\par {\bf 5BZBJ1136+1601} This BL Lac lies at $z =$ 0.574. There are five
observations contained in our catalogue taken between May and June 2014. In
all cases, the best-fit model is a {\it power law} with two absorption components
and photon index between 2.4 and 2.8. The X-ray flux in the 0.2 - 10~keV range
varies between 2.2 and 2.5$ \cdot 10^{-13}~erg~cm^{-2}~s^{-1}$. There is no
significant variability within the timescale of the observations for the default 500 s bins used.

\par {\bf 5BZBJ1210+3929} The redshift of this source is $z =$ 0.617. We
report nine observations in total from 2000 (2x), 2011 (3x), and 2012 (4x). All
observations are best fitted by a {\it power law} with two absorption components. The
photon index varies between 2.1 and 2.3 and the X-ray flux varies in the range $3.0-8.2
\cdot 10^{-12}~erg~cm^{-2}~s^{-1}$ in the 0.2 - 10 keV range. The source does
not display significant source variability, being consistent with a constant
source over the course of the observations.

\par {\bf 5BZBJ2258-3644} Of unknown redshift, two observations are present in
the catalogue, from 2005 and 2015. In both cases the spectra are best fitted by
a {\it power law} with two absorption components. The photon index of the 2005
observation is 2.9 and the flux $3.67 \cdot 10^{-13}~erg~cm^{-2}~s^{-1}$ in
the 0.2 - 10 keV range; while in the 2015 the photon index is 3.33 and the
flux $4.96 \cdot 10^{-13}~erg~cm^{-2}~s^{-1}$ in the 0.2 - 10 keV range.



\end{document}